\documentclass{aa}  

\usepackage{graphicx}
\usepackage{xcolor}
\usepackage{times}
\usepackage[varg]{txfonts}
\usepackage{amsmath}
\usepackage{graphicx}
\usepackage{amssymb}
\usepackage{natbib}
\usepackage{babel}
\usepackage{xspace}
\usepackage{array}
\usepackage{multirow}
\bibpunct{(}{)}{;}{a}{}{,}

\newcommand{\mc}[3]{\multicolumn{#1}{#2}{#3}}

\newcommand{\mrm}{\mathrm}

\newcommand {\be} {\begin {equation}}
\newcommand {\ee} {\end {equation}}
\defcitealias{khea12}{Paper~I}
\defcitealias{mwscat}{Paper~II}
\defcitealias{mwscnew}{Paper~III}
\defcitealias{mwscnew2}{Paper~IV}
\defcitealias{mwscint}{Paper~V}
\defcitealias{mwscage}{Paper~VI}
\renewcommand{\d}{\mbox{\rm d}}
\newcommand{\p}{\partial}
\newcommand{\e}{\mbox{\rm e}}

\newcommand{\msun}{\mathrm{M}_\odot}
\newcommand{\ben}{\begin{eqnarray}}
\newcommand{\een}{\end{eqnarray}}

\newcommand {\vgap} {\noalign{\vspace{0.5mm}}}

\newcommand{\GG}[1]{}

\begin{document}

\title{Global survey of star clusters in the Milky Way}

\subtitle{VII. Tidal parameters and mass function}

\author{A.~Just\inst{1}\fnmsep\thanks{corresponding author: just@ari.uni-heidelberg.de} \and
        A.E.~Piskunov\inst{2}     \and
        J.H.~Klos         \inst{1} \and
        D.A.~Kovaleva    \inst{2} \and
        E.V.~Polyachenko \inst{2} 
        }

\institute{
Zentrum f\"ur Astronomie der Universit\"at
Heidelberg, Astronomisches Rechen-Institut, M\"{o}nchhofstra\ss{}e 12-14, 69120 Heidelberg, Germany
\and
Institute of Astronomy of the Russian Acad. Sci., 48 Pyatnitskaya Str., 109017 Moscow, Russia
}

\date{Received ... / Accepted }

\abstract 
{} 
{We built Galactic open star cluster mass functions (CMFs) for different age sub-samples and spatial locations in the wider solar neighbourhood. Here, we present a simple cluster formation and evolution model to reproduce the main features of the CMFs.}
{We used an unbiased working sample 
of 2227 clusters
of the Milky Way Star Cluster (MWSC) catalogue, which occupy the heliocentric cylinders with magnitude-dependent completeness radii of 1--5~kpc. The MWSC survey provides an extended set of open star cluster parameters, including tidal radii, distances, and ages. 
From an analytic three-component Galaxy model, we derived tidal masses of clusters with a typical accuracy of about 70\%.
Our simple model includes a two-section cluster initial mass function, constant cluster formation rate, supervirial phase after a sudden expulsion of the remaining gas, and cluster mass loss due to stellar evolution and the clusters' gradual destruction in the Galactic tidal field. The dynamical evolution model is based on previous N-body simulations.}
{The obtained tidal masses have been added to the MWSC catalogue. A general CMF (GCMF), built for all cluster ages around the Sun, has a bell-like shape and extends over four decades in mass. The high-mass slope found for tidal mass $\log m_\mrm{t}/\msun \ge 2.3$ is equal to 1.14$\pm$0.07. The CMFs for different age groups show the same high-mass slopes, while the low-mass slope is nearly flat for the youngest sub-sample  (clusters younger than 20 Myr) and about $-$0.7 for the others. The inner and outer sub-samples covering Galactocentric radii $R=4.2$--8.1~kpc and 8.9--13.5~kpc, respectively, are consistent with the GCMF, once the exponential decline of the Galactic disc density is taken into account. The model suggests star formation with low efficiency of 15--20\%, where only 10\% of stars remain bound in a cluster after gas expulsion and subsequent violent relaxation. The cluster formation rate required to reproduce the observed distributions in age and mass is about $0.4\,\msun\, \mrm{pc}^{-2}\,\mrm{Gyr}^{-1}$.}
{The obtained high-mass slope of the GCMF for the wide neighbourhood of the Sun is similar to slopes determined earlier in nearby galaxies for more luminous clusters with $\log m/\msun > 3.8$. The MWSC catalogue supports models with a low star-formation efficiency, where  90\% of stars are lost quickly after gas expulsion. The obtained cluster formation rate corresponds to open clusters'  contribution to the stellar content of the thin disc at the level of 30\%. }


\titlerunning{MWSC VII. Tidal parameters and mass function}

\maketitle

\section{Introduction}\label{sec:intro}

The mass of star clusters is one of the basic parameters of these objects that specify their birth and subsequent fate. Determination of the masses for a large cluster sample is a non-trivial task. The obvious summation of masses of individual cluster members suffers from a problem of data loss due to overlapping images of stars in the central, dense regions of clusters, clogging up the counts with the surrounding background and missing the mass below or outside the observation thresholds.
There are only a few publications devoted to the cluster mass function (CMF) for our Galaxy \citep{lala03, lamea, fuma}.

Another possible way of obtaining the mass is to use the mass-to-light ratio, which is widely applied for distant extragalactic clusters \citep{zhafa99,bikea03,fallea05,dowell08,fallea09,larss09,chan10,foue12}. This approach suffers from low sensitivity to the statistics of faint stars, since the luminosity of clusters is dominated by the brightness of several most massive stars, whereas its mass is determined by the abundance of low-mass members. 

In large surveys consisting of clusters located at different distances from the Sun, these difficulties are exacerbated by the variable observational limit of the counts, which require unreliable extrapolations. These problems can be overcome by constructing a single scale of observational parameters sensitive to the cluster's mass. We propose to use the tidal radii of star clusters, which are tightly related to their bound masses. Thanks to extensive cluster member lists becoming available in recent years, the opportunity to use these data for statistical purposes has  increased considerably.

Currently, the relevant data on cluster parameters can be queried from three general data sources. The first one is the Catalogue of Open Cluster Data \citep[COCD,][]{clucat},  based on Hipparcos/Tycho all-sky catalogues and their reduction ASCC-2.5\footnote{ftp://cdsarc.u-strasbg.fr/pub/cats/I/280B}. Secondly, an extended source of  the cluster data for population studies is the Milky Way Star Clusters survey  \citep[MWSC,][referred to as Paper~I and Paper~II]{khea12,mwscat}. It has appeared as a result of open cluster studies based on all-sky ground-based catalogues, mainly the Two Micron All Sky Survey\footnote{ftp://cdsarc.u-strasbg.fr/pub/cats/II/246} \citep[2MASS,][]{cat2MASS} and the Catalog of Positions and Proper Motions on the ICRS\footnote{ftp://cdsarc.u-strasbg.fr/pub/cats/I/317} \citep[PPMXL,][]{ppmxl}. The third catalogue is an extended compilation of  \citet{daml02}, comprising collected literature data. 
Nowadays, a new cluster parameter list created by community efforts based on the most recent space born observations of the Gaia telescope has emerged.

The first results from Gaia \citep{2016A&A...595A...1G} provided a quality leap in astrometric and photometric data. They were followed by a burst of widespread astronomical interest in star cluster studies. \cite{cantatgea18} proposed a method of automatic identification of open clusters and compiled a list of 1229 clusters based on the Tycho-Gaia Astrometric Solution (TGAS) exploration experience verified with Gaia DR2 \citep{2018A&A...616A...1G} data. It relies on an algorithm of blind search of stellar over-densities in 5D parameter space, including Galactic longitude and latitude ($l$, $b$), proper motions along right ascension and declination ($\mu_{\alpha}$, $\mu_{\delta}$), and parallax ($\varpi$). The method was modified \citep{cantatgea19} and this approach was gradually applied for extending the list, compiling the membership probability and determining cluster basic parameters \citep{cantatgand20,cantatgea20,castrogea20}. In total, the authors were able to detect in Gaia DR2 1867 objects with photometric and high-quality astrometric parameters. 
 
Simultaneously,  \citet{liupang19} and \citet{simea19} published the results of their quests in Gaia DR2. \citet{liupang19} used the 
friend-of-friend (FoF) based cluster-finding method to identify in 5D parameter space 2443 clusters, including 76 that had not been mentioned in the literature before. Almost similar figures received \citet{simea19}, who identified 2080 clusters known in the literature and 207 unknown candidates using an unsupervised machine-learning method and blind search for $|b| < 20\degr$.  \citet{ferreiraea21} surveyed 200  fields selected from Gaia DR2 in the direction of the Milky Way centre and detected 34 open cluster candidates. \citet{heea21} used Gaia DR2  at $|b|<20\degr$ and $\varpi>0.2$ mas and re-identified 2080 already known clusters and also found 74 more new open cluster candidates. \cite{heea22} investigated the Gaia EDR3 \citep{2021A&A...649A...1G} data from the point of Galactic latitudes higher than $|b|>20\degr$ and close to the Sun ($\varpi>0.8$ mas) clusters. In this area, they found 886 objects, with 270 that had not been catalogued before, and 46 of the latter reside at relatively high latitudes.
Recently \citet{haoea22} were hunting for star clusters in Gaia EDR3. At Galactic latitudes of $|b|\leq 20\degr,$ they were able to find 1930 previously known open clusters, 82 known globular clusters, and 704 new open cluster candidates. 

Most of the above results were derived with the help of the Density-Based Spatial Clustering of Applications with Noise \citep[DBSCAN, introduced in][]{esterea96}, one of two clustering algorithms that can operate in the presence of background contamination typical to the star cluster environment. More recently, \citet{huntreff21} compared the effectiveness of three scanning approaches based on the Gaia DR2 catalogue in 100 selected sky areas located along the Milky Way. They found that the second known algorithm Hierarchical DBSCAN \citep[HDBSCAN,][]{campelloea13} is able to rediscover 82\% of real clusters compared to the DBSCAN rate of 50--62\%. For the third popular algorithm, the Gaussian mixture model \citep[GMM,][]{pedregosaea_11}, they found a success rate of 33\%. As a by-product, they were also able to identify 41 new clusters. 

The flow of newly identified clusters is accompanied by newly determined Gaia-based cluster parameters (first astrometric and then photometric ones), \citet{bossiniea19} for Gaia DR2, as well as \cite{castrogea21} for Gaia EDR3, which extend the horizons for cluster population studies.  For example, \citet{soubirea18,soubirea19} used Gaia DR2-based 6D data for 861 clusters and 5~newly identified cluster pairs to study Galactic disc kinematics. 
\citet{castrogea21} used 264 open clusters younger than 80~Myr and 84 star-forming regions younger than 30~Myr, all residing around the four spiral arms nearest to the Sun to outline the arms, characterise them, and determine their kinematics. \citet{tarricqea21} studied the kinematics of the open cluster population from data of Gaia DR2 with the primary aim to investigate its kinematics and orbital properties with age. They also investigated the rotation curve of the Galactic disc traced by open clusters. \citet{andersea21} build an age distribution for 834 Gaia DR2 clusters residing within 2~kpc from the Sun. Their statistics indicate that the present cluster formation rate is 0.55~Myr$^{-1}$ kpc$^{-2}$, and only 16\% of field stars were formed in bound clusters.

We note that Gaia provides an exceptionally valuable opportunity to study the star cluster outskirts, which had hitherto escaped the community's attention. However, after reporting the impressive results of studies of nearby clusters \citep{roeserea19,roesersch19}, interest has strongly risen.   \citet{angeloea20,angeloea21} used Gaia DR2 to study a number of low-contrast clusters. Based on safe membership,  for 65~clusters, they determined the structure and dynamical parameters (King and half-mass radii, ages, and crossing times) with the purpose of estimating the cluster dynamical state.  \citet{tarricqea22} using advanced tools (HDBSCAN, GMM algorithms) for Gaia EDR3 data have established wide-area memberships (within 50~pc) of 467 clusters. For 389 of them, they were able to construct radial profiles, fit to them King profiles (146 quality fits) and determined tidal parameters. For 71 clusters they identified the tidal tails and studied the degree of mass segregation. \citet{zhongea22} continued to exploit this approach with Gaia EDR3 and studied wide area radial profiles of 256 open clusters within 1--2~kpc from the Sun. As they find, the profiles can be represented as a sum of two distributions: the internal profile, which follows a King model, and the outer one obeying a log-normal law. 

As it follows from above, the current status of cluster population characterization with Gaia is unstable and swiftly developing. Despite the high quality and homogeneity of data on individual stars, this cannot be determined for a representative cluster sample nor for high-significance astrophysical parameters, such as the luminosities or masses characterising the cluster population with regard to its formation and evolution. This is why we base this study on the MWSC sample, as it provides the necessary fundamental qualities such as estimated completeness and an extended set of relevant parameters, including almost 100\%-coverage with tidal radii. Our ultimate aim is based on independent and uniform estimates of cluster masses drawn in this study from tidal radii. We strive to build a Galactic star cluster mass function, study its temporal and spatial variations, and  advance a cluster population model which adequately describes these features.
\begin{figure*}[hbt]
\centering
 \includegraphics[width=0.3275\hsize,clip=]{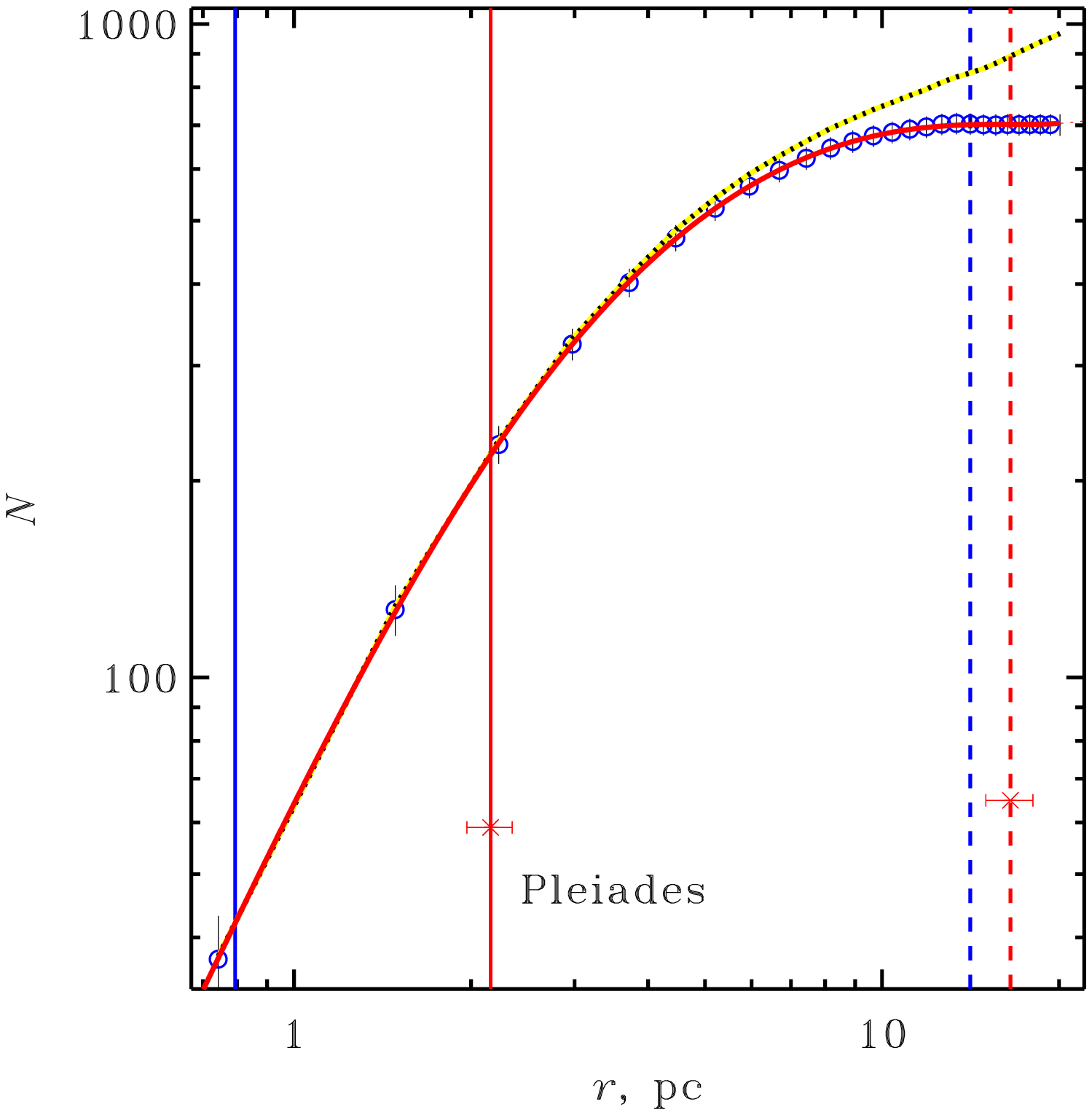}
 \includegraphics[width=0.2975\hsize,clip=]{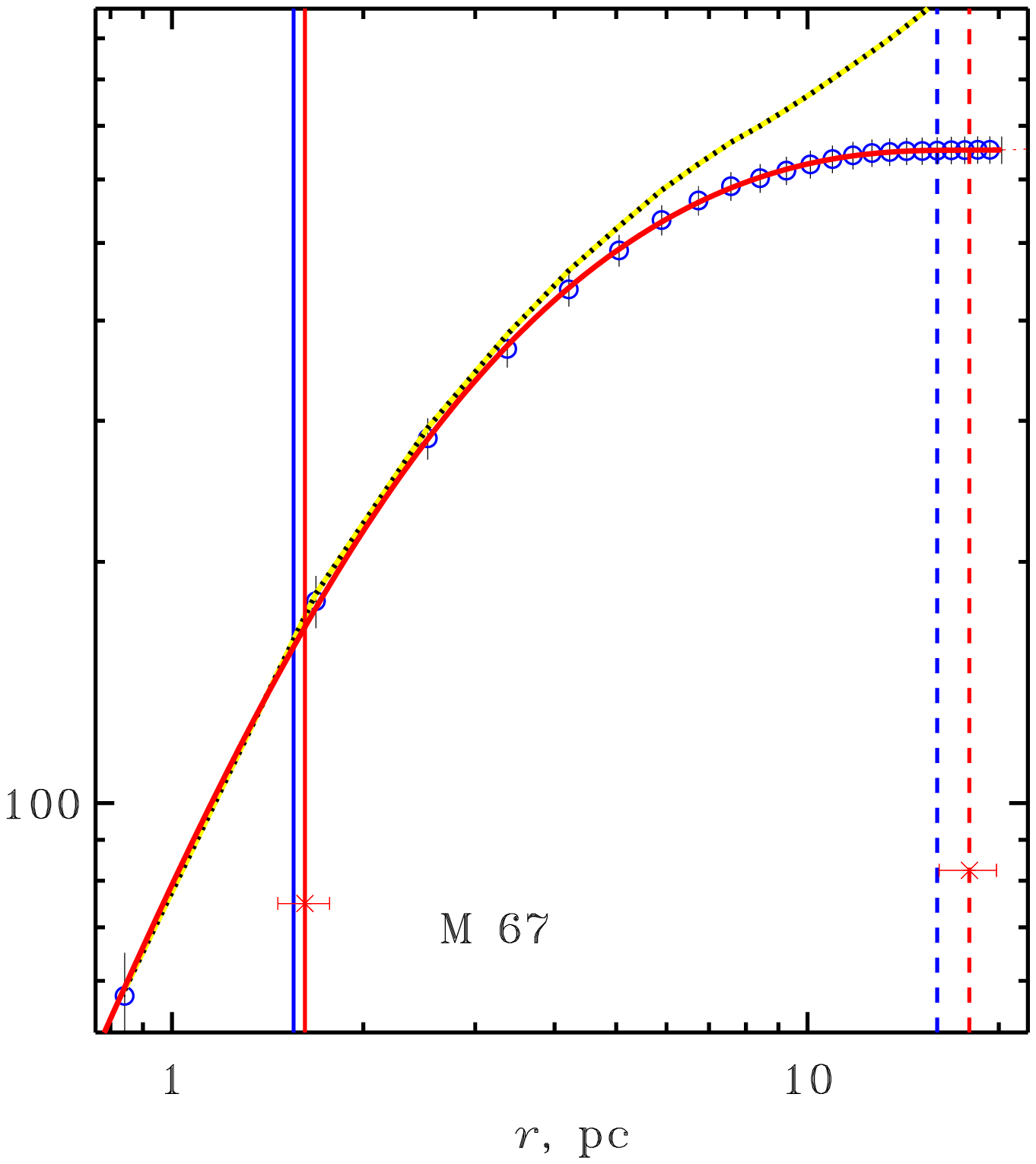}
 \includegraphics[width=0.29\hsize,clip=]{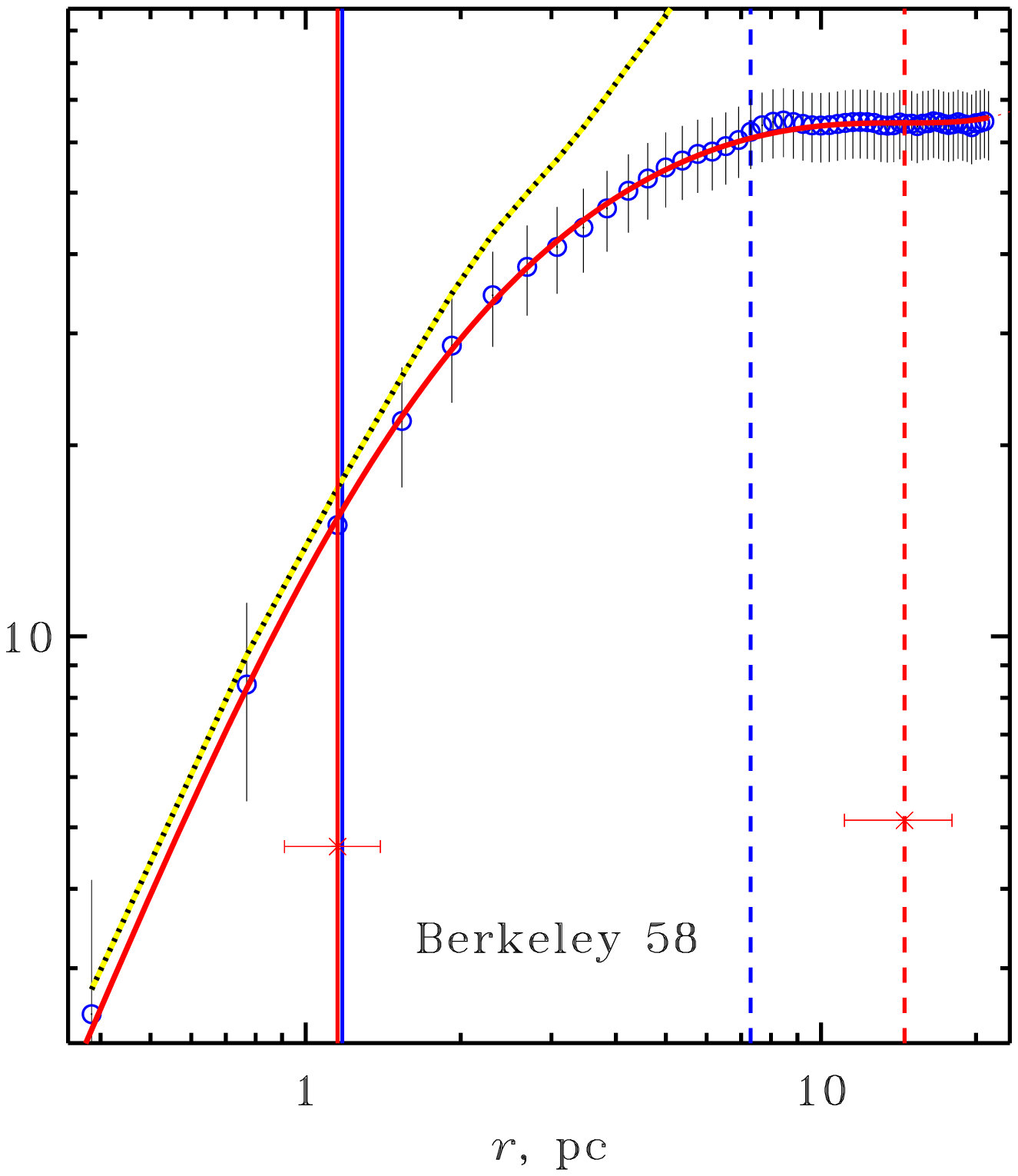}
\caption{Fit of King curves to observed density profiles of open clusters residing at different heliocentric distances: small (Pleiades), medium (M67), and large (Berkley~58), shown from left to right. The curves represent the observed distribution of safe cluster members (dots), the one corrected for the residual background (circles with Poisson error bars), and the fitted King profile (solid red curve). Vertical lines indicate core radii ($r_\textrm{c}$, solid) and tidal radii ($r_\textrm{t}$, dashed). Blue lines indicate visual estimates, made earlier in MWSC, red ones are the current values of $r_\textrm{c}$ and $r_\textrm{t}$  derived from the profile fit. Horizontal bars indicate fit errors.
}\label{fig:fit3}
\end{figure*}

In Sect.~2,  we briefly outline the cluster data set applied in this study. In Sect.~3, we characterise the used tidal radii and in Sect.~4, we describe how they were converted to \textbf{}tidal masses. Section 5 is devoted to the construction of the general cluster mass function (GCMF). We pay special attention to the data completeness, the unbiased sample construction issue, and describe the technique of the GCMF construction and define the basic relevant entities. We investigate the stability of the results against some accompanying effects and compare the results with recent literature data. Sect.~6 is devoted to the issues of temporal and spatial variations of the CMF, including the cluster initial mass runction (CIMF), and evolutionary changes of the CMF. In Sect.~7, we propose a simple model of the formation and evolution of the Galactic disc cluster population, which can reproduce the GCMF and the main features of the CMFs of different age groups. In Sect.~8, we give our  results and conclusions.

\section{Data}\label{sec:data}

For this study, we use the results of our all-sky MWSC survey,  aimed at a homogeneous, extensive, and multilateral characterisation of Galactic open clusters based on the analysis of their spatial, kinematic, and photometric cluster membership. It has been made possible thanks to accurate and systematic all-sky ground-based photometry and astrometry published at the beginning of the 2000s in the 2MASS \citep{cat2MASS} and PPMXL \citep{ppmxl} catalogues. The complete list of  Galactic star clusters known at that time was examined and additional effort was undertaken to search for as-yet unidentified objects.

\begin{figure*}[t!]
\centering
 \includegraphics[width=0.85\hsize,clip=]{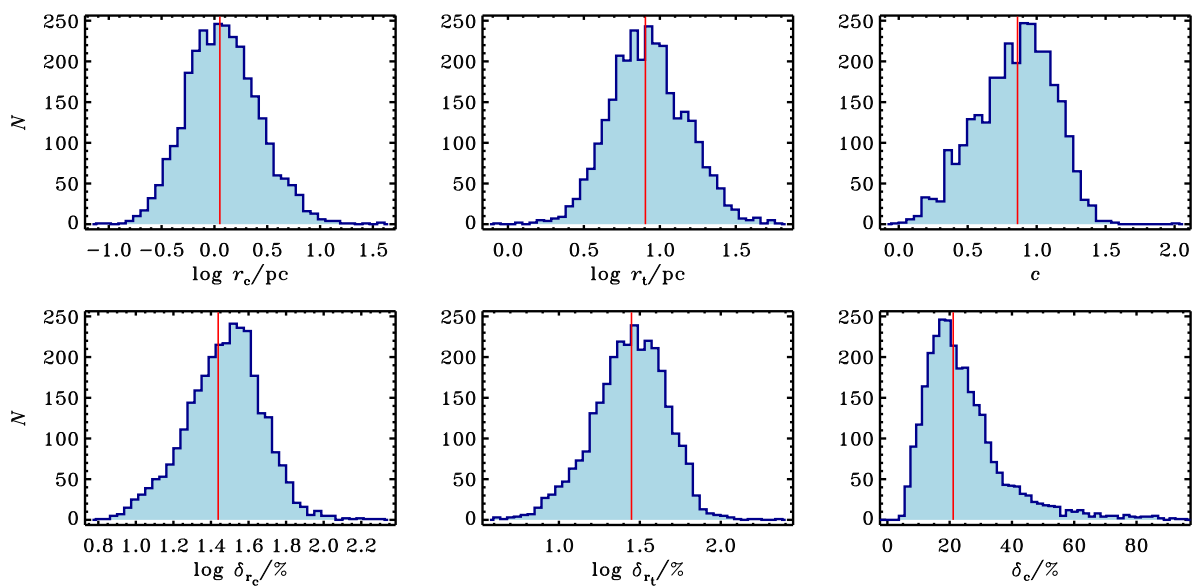}
\caption{Tidal parameters of MWSC clusters. Distributions of $\log r_\textrm{c}$, $\log r_\textrm{t}$, and $c$ (top row) and of their relative errors (bottom row). Vertical lines show sample averages. 
}\label{fig:rct_dstr}
\end{figure*}

The sky area covered to study a particular cluster is limited only by the properties of the surrounding stellar field and usually is equal to the cluster size taken from the literature, along with an additional ring of $0.3\degr$ width if no indication of further area extensions was found in the study. The magnitude depth of MWSC counts is as a rule equal to $K_\textrm{s} =$ 15--16 mag. Over 63 million stars in these areas were considered and about 0.2\% were identified as the most probable cluster members. It should be noted that the depth exceeds the completeness limit, which depends on the local conditions and varies with cluster area around $K_\textrm{s} =$ 13--14 mag.

The selected members were used for the cluster parameter determination. In every cluster, the spatial distribution of cluster members was built. The member density profiles were built for every cluster, and angular structure parameters were determined. These were used as initial conditions to establish a standardised scale of cluster structure parameters via the fit of a three-parameter King model to the observed density profiles. This gave us a unique scale of cluster structure parameters for more than 3000 MW open clusters. 
The $(K_\textrm{s},\,J-K_\textrm{s})$, $(K_\textrm{s},\,J-H)$ CMDs and two-colour $(H-K_\textrm{s},\,J-H)$, and  $(Q_{JHK_\textrm{s}},\,J-K_\textrm{s})$ diagrams were constructed and used for reddening and photometric distance determination \citepalias[for more details see][]{khea12}. These parameters were used to fit theoretical isochrones to the cluster member CMDs and to determine the cluster age, reddening, and distance modulus. For the age indication, both the turn-off and turn-on (if observed) CMD regions (where the isochrone respectively departs from and joins ZAMS) were used. The integrated near-infrared (NIR) magnitudes and colours of MWSC clusters are also additionally computed. 

The results of the MWSC were published in a series of papers and submitted as online catalogues to the CDS archive\footnote{https://cdsarc.cds.unistra.fr/ftp/J/A+A/558/A53;\\ https://cdsarc.cds.unistra.fr/ftp/J/A+A/568/A51; https://cdsarc.cds.unistra.fr/ftp/J/A+A/581/A39; https://cdsarc.cds.unistra.fr/ftp/J/A+A/585/A101.}. In the above-mentioned work of \citetalias{khea12}, the cluster membership pipeline is described, which is the main instrument of the MWSC. It provides data on individual stars observed in the search area  in the form of combined astrometric and photometric cluster membership probabilities and integrated parameters of the derived member ensembles characterising the clusters themselves. For all studied clusters, the MWSC provided an extended list of their parameters: structural ones, including the sizes of cluster core, total cluster extent, cluster proper motions, and, when possible, radial velocities, as well as photometric parameters: reddening, distance modulus, and age. All these data were presented in \citetalias{khea12} for the second quadrant objects and in \citetalias{mwscat}, for the rest of the known Galactic clusters. In addition, we have searched for  as-yet unidentified clusters \citep[][hereafter Papers~III and IV]{mwscnew,mwscnew2}. In  \citetalias{mwscnew} the new candidates were searched as sky field over-densities for apparent Colour-Magnitude Diagrams (CMD) reproducing patterns typical to older clusters, while in \citetalias{mwscnew2} the new candidates were searched as enhanced densities in the proper-motion vector-point diagrams (VPD). At last, the integrated near-infrared 2MASS magnitudes and colours of MWSC clusters were also additionally computed \citep[][hereafter, Paper~V]{mwscint}. 

As a result, our final compilation contains 3210 previously known and newly identified star clusters, with 3063 of them classified as open clusters. The MWSC covers a substantial part of the Galactic disc radius ($R$)  from the very centre at $R\simeq 2$ kpc to the outer regions at $R\simeq 20$ kpc. An important feature of the MWSC is the uniformity of the parameters obtained during its execution. This homogeneity is ensured by the fact that the entire survey is performed by one person (N.V. Kharchenko) in the framework of uniform methods and approaches. The sample of clusters built in such a way contains an unprecedentedly wide set of various parameters of a spatial, kinematic, photometric, and astrophysical nature, representing a perfect tool for studying the Galactic star cluster population. It was already used to study the open cluster near-infrared luminosity functions \citepalias{mwscint} and the history of the cluster formation rate \citep[][hereafter Paper~VI]{mwscage}. The next natural step of its application is to study various issues related to the star cluster mass function.

\section{Tidal radii}\label{sec:tira}

The sizes of the star clusters have a clear physical meaning, they characterise the mass and population of the grouping and its dynamic state. An accurate and homogeneous fixation of the boundaries of open star clusters is a non-trivial task: low star density in the cluster outskirts, an admixture of residual field stars, the expected non-sphericity of the clusters, the possible presence of secondary structures (e.g. tidal tails), variations of observations, and dependence on observation sources -- all of these aspects require a certain standard approach for the cluster size definition.

The method we used is based on the well-known empirical King’s model \citep{king62}, which describes the radial profiles of the surface density of stars observed in globular clusters using curves depending on the parameters, $r_{\rm c}$, $r_{\rm t}$, and $k$. According to King's definition, $r_{\rm c}$ is the radius of the core, $r_{\rm t}$ is the tidal radius, and $k$ is the normalising factor associated with the central density of the cluster. The logarithm of the radii ratio $c=\log (r_{\rm t}/r_{\rm c})$,
called concentration, is also widely used. In addition to globular clusters, this approach was also used to determine the King parameters of several nearby open clusters \citep[see e.g. earlier][]{ramer98a,ramer98b} or recently \citet{,angeloea20,angeloea21,tarricqea22}. However, the direct definition of differential density profiles does not work for most open clusters due to the relatively small number of members observed in them and the relatively low density at the cluster's periphery. This causes a systematic distortion of the outer areas of the clusters, especially important for determining tidal radii. Therefore, to reduce the effect of poor statistics, we used the integral form of the King’s profile for our purposes:
\be\label{eq:kingi}
 \begin{split}
  n(r) = \pi\, r_{\rm c}^{2}k\, 
          &\left\{\ln[1+(r/r_{\rm c})^{2}]-4\frac{\left[1+(r/r_{\rm c})^{2}\right]^{1/2}-1}{\left[1+(r_{\rm t}/r_{\rm c})^{2}\right]^{1/2}} \right.\\
          &\left.+\frac{(r/r_{\rm c})^{2}}{1+(r_{\rm t}/r_{\rm c})^{2}}\right\}\,,
 \end{split}
\ee
where $n(r)$ is the number of cluster members within a sky area with radius ($r$). In COCD, we fit integrated profiles of 236 clusters \citep{clumart}.

To construct the observed cluster profile, we counted cluster members in concentric rings around the centre of the cluster at distances up to $ 5r_2 $, where $ r_2 $ is the apparent radius of the cluster's corona, obtained from MWSC. To standardise the density profiles observed for clusters embedded in different environments, we used the most reliable members of the cluster that provide the exhaustive completeness of the data and help to avoid the respective bias in the working sample. Special attention was given to the cluster membership probability and apparent magnitude limits. Their values were varied in a broad range and optimised by the goodness-of-fit parameter. The sampling parameters were selected individually for each cluster and varied depending on its individual properties: distances, degrees of immersion in the surrounding stellar back- and foreground, uniformity, and heaviness of interstellar extinction. The decisive factor was the completeness and purity of the data, especially in the outer parts of the cluster. Therefore, we took additional measures to take into account the residual background and remove “false” member field stars that satisfy the member selection filters (for example, co-moving field stars on the proper-motion diagram or stars projecting onto the cluster main sequence in the colour-magnitude diagram). To do this, we calculated the level of residual density of field stars, typical for the surrounding stellar field, and subtracted it from the primary empirical profile. The approximation and determination of the parameters were carried out using the {\tt MPFIT} procedure from the \cite{markwdt09} {\tt IDL} library.

\begin{figure}[tb]
\centering
\includegraphics[width=0.90\hsize,clip=]{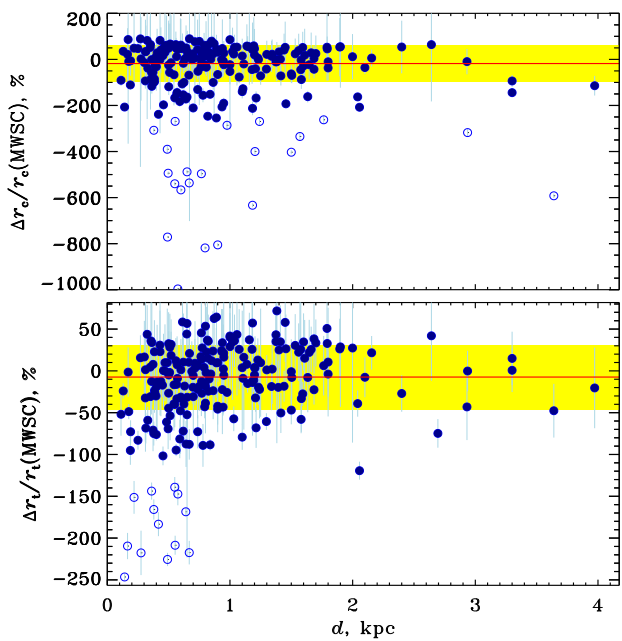}
\caption{Relative difference in COCD and MWSC estimates of $r_{\rm c}$ and $r_{\rm t}$ radii as function of MWSC heliocentric distance ($d$). Red lines show average differences and yellow stripes are respective standard deviations. Statistical parameters were computed with the help of data marked with filled circles, open circles are omitted as 3-$\sigma$ outliers (24 and 13 for $r_{\rm c}$ and $r_{\rm t}$, respectively). Vertical bars are fit errors.}\label{fig:rct_cmp}
\end{figure}

Figure \ref{fig:fit3} shows the empirical profile approximation for clusters residing at different heliocentric distances: 0.13 kpc for one of the closest to the Sun open cluster Pleiades, 0.89 kpc for a typical to MWSC sample cluster M67, and 2.7 kpc for a remote cluster Berkley~58. It can be seen that the approximation works equally well at different distances and that the profile correction for residual background plays a critical role in the correct determination of tidal radii.

Using the described method, we determined the tidal parameters of 3017 open clusters (98\% of all MWSC objects classified as open clusters). For the remaining 2\%, satisfactory density profiles were not built due to the poor quality of the input data (poor statistics, ragged absorption, near bright star, etc.). The typical (most frequent) determination accuracy is 29 to 33\% for the radii. The distribution of these parameters and their accuracy for the clusters studied is shown in Fig.~\ref{fig:rct_dstr}. It can be seen that, according to our determination, the typical core radius of open clusters is 1.1 pc, which corresponds to widespread belief about the sizes of open clusters. The actual size of a typical open cluster confined by the gravitational field of the Galaxy ($ r_{\rm t} $) is, by our determination, 8.4 pc, and the typical concentration, $c,$ of disc clusters is 0.91.

Figure \ref{fig:rct_cmp} compares our determinations of tidal parameters made based on data for 222 common clusters of COCD and MWSC depending on MWSC heliocentric distances. Methods for determining the radii of both samples differ only in details related to the specifics of the data, in particular, with a brighter completeness magnitude in COCD. Despite the difference in observational properties, both surveys show good agreement: the core radii differ by no more than 18\% and the tidal radii by 7\% on average. At the same time, there are no systematic differences up to the distances of about 3~kpc. Some trend over larger distances appears to be associated with lower reliability of structure data in remote clusters in COCD. 

\section{Tidal masses}\label{sec:tima}

To determine the tidal mass of the cluster ($m_{\rm t}$), we follow \citet{king62} and use a condition for the balance of gravitational forces between the Galaxy and the cluster on a circular orbit:
\be
m_{\rm t} = \frac{r_{\rm J}^3}{G}\left(\frac{1}{R}\frac{\partial \Phi}{\partial R} - \frac{\partial^2\Phi}{\partial R^2}\right).\label{eq:rjgen}
\ee
Here, $r_{\rm J}$ is the Jacobi radius (distance from the cluster centre to Lagrange points, $L_1,$ or $L_2$, where its self-gravity is equal to the Galaxy field in the corotating frame), $\Phi$ is the Milky Way potential, $R$ is the Galactocentric radius, $G$ is the gravitational constant. 

For an explicit representation of the Galactic potential, we use a Plummer-Kuzmin three-component model \citep{miyana75}:
\be
\Phi_i(R,z)=-\frac{G\,M_i}{\sqrt{R^2+\left(a_i+\sqrt{b_{i}^2+z^2}\right)^2}}\,,
\ee
where $z$ is the distance from the Galactic plane, $i=b,d,h$ are indices of the components bulge, disc, and halo respectively, and $M_i$ are the masses of these components. Model parameters $M_i$, $a_i$, and $b_i$ were optimised to represent the Galactic rotation curve adopting $R_{\rm \odot}=8.5$\,kpc for the Galactocentric distance of the Sun \citep{elli}. These are shown in Table~\ref{tab:pargp}.

\begin{table}[b]
 \begin{center}
 \caption{Parameters of Galactic potential components.}\label{tab:pargp}
\begin{tabular}[c]{cccc}
\hline
\hline
\vgap
Component& $M,\,\msun$   & $a$, kpc& $b$, kpc    \\
\vgap
\hline
\vgap
bulge    & $1.4\times 10^{10}$ & 0.0     & 0.3\\
disc     & $9.0\times 10^{10}$ & 3.3     & 0.3\\
halo     & $7.0\times 10^{11}$ & 0.0     & 25.0\\
\vgap
\hline 
\end{tabular}
\end{center}
\end{table}

\begin{figure}[t!]
\centering
\includegraphics[width=0.90\hsize]{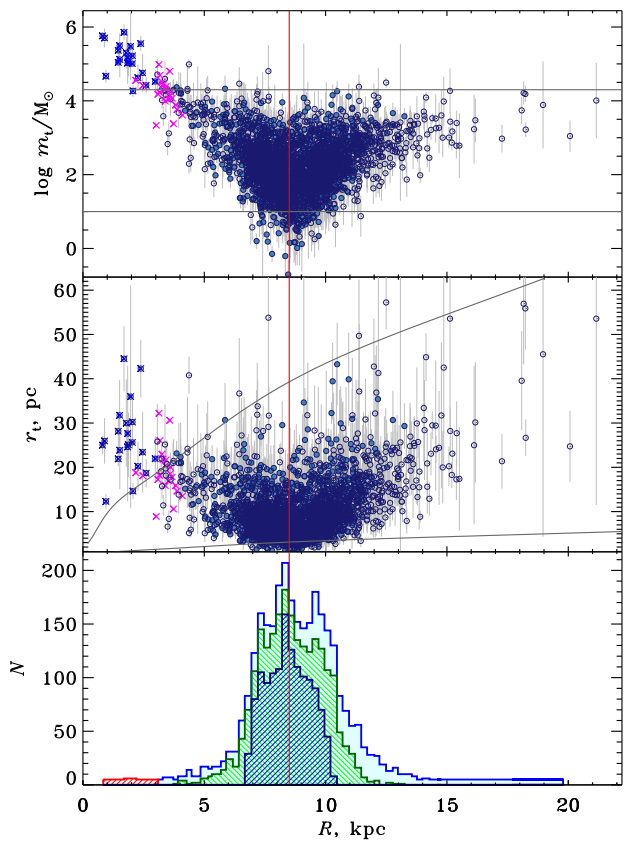}
 \caption{Radial distributions of tidal masses (top panel), radii (middle panel), and cluster counts (bottom panel). Filled and open circles in the upper two panels show clusters inside and outside the magnitude-dependent completeness limits described in Sect.~\ref{sec:datcompl}, respectively. Bars are errors in the determination of the tidal parameters. Crosses are central ($R\le 3$ kpc) clusters before (blue) and after (magenta) the artificial decrease of $0.6$ mag in their distance moduli is applied. Solid lines are theoretical relations for cases of a constant mass of $m_{\rm t}=2\times 10^4\,\msun$ (upper line) and $m_{\rm t}=10\,\msun$ (lower line). The red vertical line shows the Galactocentric radius of the Sun. Histograms in the bottom panel show raw counts for all clusters (background), ones within the single completeness limit (foreground), and those within magnitude-dependent completeness limits (in between). The red histogram shows the central clusters.}\label{fig:rmt_rg}
\end{figure}

At the distance ($R$) from the Galactic centre, the angular velocity ($\Omega$) and the epicyclic frequency ($\kappa$) can be expressed as follows:
\begin{flalign}
\Omega^2&=\frac{1}{R}\frac{\partial \Phi}{\partial R}\,,\\
\kappa^2   &=\frac{\partial^2\Phi}{\partial R^2}+\frac{3}{R}\frac{\partial\Phi}{\partial R} = 
4\Omega^2 + R \frac{\d\Omega^2}{\d R}\,,
\end{flalign}
and Eq. \ref{eq:rjgen} is written as:
\be\label{eq:jamas}
m_{\rm t}=\frac{r_{\rm J}^3}{G}(4-\beta^2)\Omega^2,
\ee
where $\Omega$ and $\beta \equiv \kappa/\Omega$ depend on $R$. 

To apply Eq.~\ref{eq:jamas}, in practice, it is necessary to know its main components in advance. For example, the Jacobi radius $r_{\rm J}$ could be scaled via the tidal radius computed from the fit of King profiles to the observed radial distributions of cluster member counts. In turn, the parameters $\beta$ and $\Omega$ can be taken from the Galactic potential model. 

The relation between Jacobi and tidal radii for realistic clusters in the framework of the adopted Galactic potential model was studied using N-body calculations by \cite{ernstea10}. They show that the ratio, $r_{\rm t}/r_{\rm J}$, depends on cluster coordinates in the sky and, to a lower degree, on their age. For Galactic latitudes,  $|b|<30^\circ$, where the bulk of open clusters reside, this ratio varies between 1.00 and 1.20 with the most frequent value of 1.06. A maximum of about 1.4 is reached near the Galactic poles. Although the derived bias is well within the typical random error of the $r_{\rm t}$ determination (see Fig.\ref{fig:rct_dstr}), it leads to a perceptible change in cluster masses. The respective ratio of tidal and Jacobi masses varies for different latitudes and ages between 1.0 and 2.0, having a broad asymmetric maximum at about 1.2 \citep[see Fig.~8 in] []{ernstea10}. Having in mind these figures, we adopt hereafter as zero-order approximation a hypothesis on the equality of tidal and Jacobi radii.    

Figure~\ref{fig:rmt_rg} shows the distribution of clusters with respect to their Galactocentric distance ($R$). The dependencies of the derived tidal masses and radii of clusters on $R$ are given in the top and middle panels, respectively. It can be seen that the constructed relations have several characteristic details. Despite the large extent of massive clusters, the observed clusters occupy only part of the MWSC sample space. Both relations have clearly defined boundaries of $R$, reducing its width with decreasing $r_{\rm t}$ or $m_{\rm t}$, respectively. This behaviour indicates the magnitude limit of the sample (which is the case for the MWSC survey) and the implicit dependence of the considered parameters on the cluster brightness, which is the basic factor affecting cluster visibility. 

Solid lines in the top and middle panels of Fig.~\ref{fig:rmt_rg} show the upper and lower bounds of the samples. As can be seen, the lower boundary of the observed tidal masses is about or somewhat less than $10\,\msun$, then, as the upper limit is $m_{\rm t}=2\times 10^4\,\msun$ and practically does not depend on $R$. An exception is the innermost clusters (see below). 
The distribution of the tidal radii of clusters is skewed along $R$: the distribution boundaries drawn on the middle panel are calculated by Eq.~\ref{eq:jamas} for cases of constant mass $10\,\msun$ and $2\times 10^4\,\msun$. As follows from the middle panel, relatively extended clusters should be observed in the outer areas of the disc and more compact ones in the centre.

Filled circles show
the sub-sample of clusters inside the magnitude-dependent completeness limit (derived in Sect. ~\ref{sec:datcompl}), which is free from selection effects. Clusters outside it may also be useful for studying the properties of their population.
As follows from the top panel of Fig.~\ref{fig:rmt_rg}, the completeness limits are mass-dependent. For massive clusters with $\log m_{\rm t}/\msun=4$, the variation in $R$ reaches 10~kpc, while for low-mass clusters with $\log m_{\rm t}/\msun=1$ it is as low as 2~kpc. Thus, our sample is biased and must be corrected to study the population characteristics (for example, the mass function).

The upper mass limit is broken by 21 clusters located in the very centre of the Milky Way ($R\le 3$ kpc), having about the same size as external clusters. Their resultant masses are much larger (by 1.5 orders of magnitude) than the total upper mass limit noted above. Of the two possibilities that explain this peculiarity: the real difference between the central clusters and the whole population, or the effect of systematic error we prefer the second one. We consider it quite likely that the distances to some clusters observed in the direction of the centre of the Milky Way are overestimated in the MWSC. This may be due to the ragged structure of dense dust clouds in this region of the sky: 15 of 21  `central' clusters reside at $|b|<6\degr$, and five more at $|b|=6-15\degr$. Additionally, these clusters are located near the observed MWSC magnitude limit. Therefore, only branches of red giants that are very unreliable distance indicators are available for observation. To estimate the effect of distance, we artificially reduced the modulus of the distance of these clusters by $\Delta ({K}_{\rm S}-M_{K_{\rm S}})=0.6$ mag. The result is shown in both diagrams in Fig.~\ref{fig:rmt_rg} with crosses. As can be seen in this case, the central clusters occupy a natural position on the radial diagrams typical to the rest of the clusters.
The bottom panel of Fig.~\ref{fig:rmt_rg} provides the cluster counts as a function of the Galactocentric radius for the different samples.

\begin{figure}[t!]
   \centering
\includegraphics[width=0.90\hsize,clip=]{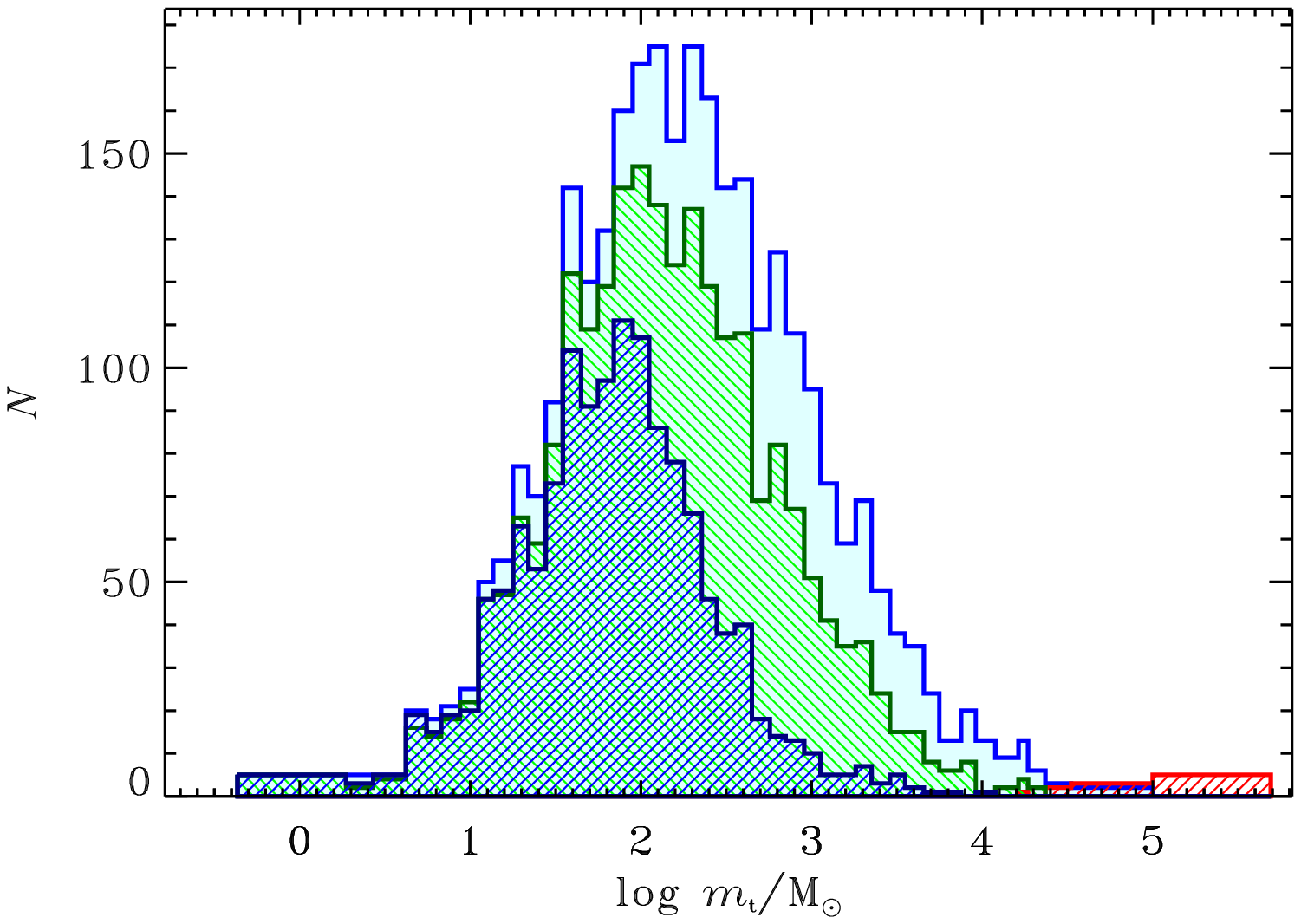}
\includegraphics[width=0.90\hsize,clip=]{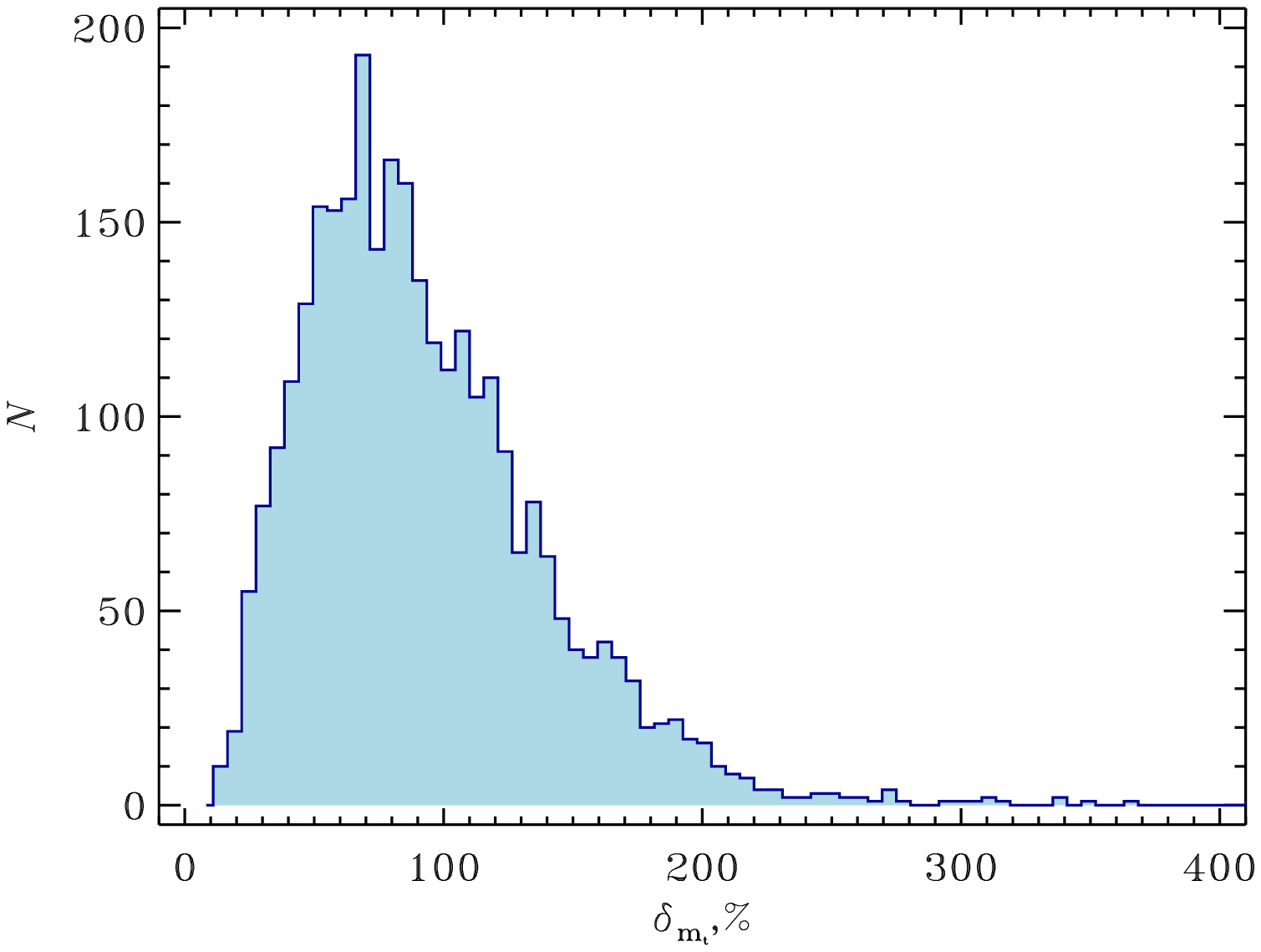}
\caption{Raw distributions of MWSC clusters with tidal mass (top panel) and with its relative error (bottom panel). Different histograms in the top panel correspond to different cluster samples. The total sample is shown in the background with a solid cyan histogram, the clusters selected within the single completeness circle are shown with a histogram hatched with blue (foreground) and those within their magnitude-dependent completeness limits are shown with a back-hatched green histogram (in between). The red-colour massive end of the distribution is built of `central' ($R \le 3$ kpc) clusters. 
}
\label{fig:his_nlgm}
\end{figure}

\section{Cluster mass function}\label{sec:clumf}

The raw cluster mass distribution of 3017 MWSC clusters with $m_{\rm t}$ determinations is shown in Fig.~\ref{fig:his_nlgm} as a background histogram. One can see that it is dominated by small/medium-mass clusters with  $\log m/\mrm{M}_\sun=$ 1.7--2.7. After taking into account the data completeness effect it is used as a basis for the mass function construction. 

\subsection{Data completeness}\label{sec:datcompl}

As cluster counts show, the MWSC can be classified as a magnitude-limited sample \citepalias[for details, see][]{mwscint}.  The surface density profile for such a sample can be represented schematically by a flat inner area, where the data incompleteness is negligible and by a long outer tail of gradually decreasing density, which is biased by the survey incompleteness at faint magnitudes. The incompleteness can be quantified in a statistical sense as a measure of the decrease of the observed surface density compared to the averaged local density \citep[see e.g.][]{mora13}. We note that as a measure of the distance we use a Galactic plane projection, $d_{xy}$, of solar-centric distance, $d$. The radius of the flat area, $\hat{d}_{xy}$, is then called the completeness limit of the survey. Once established, the bias-free statistic is gathered within the completeness limit.

This method (which we call hereafter the 'single' completeness limit approach) is attractive due to its simplicity and is commonly used, but it is inherently biased for objects which are absolutely fainter or brighter than the clusters typical for the given sample. For example, when applying the single-limit approach to faint objects, which can be observed near the Sun only, one underestimates their density when one divides their counts by the completeness area defined by the common completeness limit. In contrast, since the typical distance to bright objects may exceed the completeness limit, one can lose them from the statistics. Therefore, to avoid important biases which might affect the 
low- and/or high-mass extrema of the distribution,
we decided to abandon the single-limit approach. Instead, we apply a strategy used previously for the cluster luminosity function \citepalias{mwscint} and age distribution construction \citepalias{mwscage}, which collects star clusters of different absolute magnitudes from proportionally extended completeness areas. We refer to this method as a 'magnitude-dependent' completeness limit approach.

This procedure became feasible since in \citetalias{mwscint} we determined integrated NIR magnitudes for all MWSC clusters and built magnitude-dependent completeness limits. For the absolute integrated magnitude $I(M_{K_{\rm S}})$ in the $K_{\rm S}$ passband, this relation can be written for the total cluster sample as:
\be
\hat{d}_{xy} = p - q\, I(M_{K_{\rm S}})\,        \label{eq:dciksrel}
,\ee
with $p=0.36$~kpc and $q=0.54$~kpc\,mag$^{-1}$ (see Sect.~\ref{sec:cmfr} and Table~\ref{tab:tabpq} for more details). To derive  the relation coefficients at the distance scale extrema more precisely, we repeated the procedure described in \citetalias{mwscint}, using a more disturbance-resistant (sliding window average) approach.  
As in \citetalias{mwscint}, the completeness distances cover a large range up to about 5~kpc and the MWSC is generally complete within 1.8~kpc from the Sun (except for the faintest clusters). About half of all MWSC open clusters are inside this single completeness limit.

\begin{figure}[t!]
   \centering
\includegraphics[width=0.9\hsize,clip=]{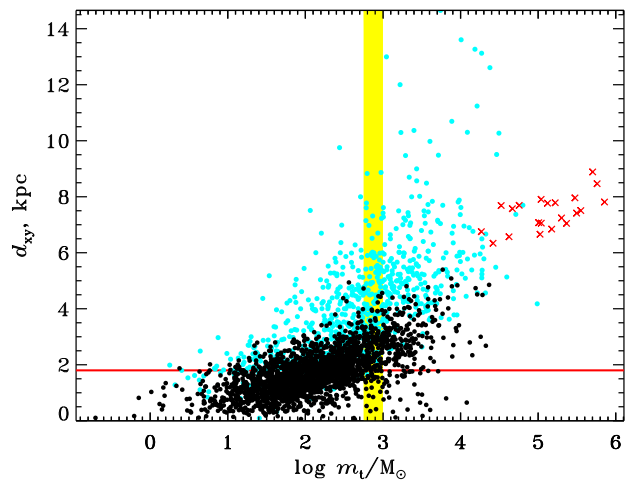}
\caption{Distribution of cluster distances with mass. The clusters located inside and outside the magnitude-dependent completeness limits computed in Eq.~\ref{eq:dciksrel} are shown with black and light blue dots, respectively. 
The single completeness limit for the MWSC sample is given by the horizontal red line. A vertical yellow stripe is given for illustration and indicates an arbitrary $\log m$-box with clusters of the two kinds falling in it. Red crosses show `central' ($R\le 3$ kpc) clusters excluded from further consideration.
}
\label{fig:dxy_lgm}
\end{figure}


In Fig.~\ref{fig:his_nlgm} (top panel), we compare the distributions of tidal masses of all clusters (except `central' ones) with known $m_{\rm t}$ (2996 objects, cyan histogram), and of the complete samples represented by hatched histograms (blue for the single completeness limit with 1328 clusters, and green for the magnitude-dependent completeness limit with 2227 clusters). One can see that the distributions coincide at the low-mass end ($\log m_\mrm{t}/\mrm{M_\sun} \le 1.5$). At higher masses, the magnitude-dependent completeness limits exceed 1.8~kpc resulting in a larger number of clusters. 

The difference between the `complete' samples can be understood with the help of Fig.~\ref{fig:dxy_lgm}, where we compare both these approaches in the completeness treatment in the `$d_{xy}\, \mrm{versus}\, \log m_\mrm{t}/\mrm{M_\sun}$' diagram. We can see that the single completeness limit approach cuts almost all potentially useful masses higher than $\log m_\mrm{t}/\mrm{M_\sun} \approx 2.7$, leaving only an insignificant number of objects for the massive end of the mass function. In the alternative case, we can extend the size of the completeness area more than by a factor of two: the upper limit of the completeness zone reaches for the intrinsically brightest clusters $d_{xy}=5$ kpc. Hereafter, we use the magnitude-dependent completeness limits to have better statistics at the high-mass end (general sample).

The red-coloured histogram in the top panel of Fig.~\ref{fig:his_nlgm} shows the innermost clusters discussed in  Sect.~\ref{sec:tima}. Following this discussion, we consider the derived masses of these clusters to be unrealistic, exclude them from further discussion, and assume that the massive end of the Galactic star cluster mass function extends to about $\log m_\mrm{t}/\mrm{M_\sun}=4.4$ only. At small masses, the raw mass distribution ends at $\log m_\mrm{t}/\mrm{M_\sun} \approx 0$. Since according to the bottom panel of Fig.~\ref{fig:his_nlgm}, the distribution of relative errors in the determination of tidal masses sharply peaks at about $\delta_{m_{\rm t}} \approx$ 70\%, corresponding to $\varepsilon(\log m_\mrm{t}/\mrm{M_\sun})\approx 0.3$, we can assume that the real low-mass end of the masses of star clusters is close to $\log m_\mrm{t}/\mrm{M_\sun}=1$, and the lower masses form the error tail at this limit.

\subsection{Construction of the mass function}\label{sec:method}

We define the cluster mass distribution $\varphi(m)$ as a surface density of objects in the unit interval of mass $m$:
\be
\varphi(m)= \frac{1}{S(m)}\,\frac{\mrm{d} N(m)}{\mrm{d} m}\,,   
\ee 
where $\mrm{d} N(m)$ is the number of clusters with masses between $m$ and $m+\mrm{d} m$ residing within the completeness area $S(m)$. It is related to the more convenient logarithmic mass distribution: 
\be
 \phi(m) = \frac{1}{S(m)}\,\frac{\mrm{d}N(m)}{\mrm{d}\log m}  \label{eq:massdistr}
,\ee
via
\be
 \varphi(m) = \frac{\log \e}{m}\,\phi(m).  \label{eq:varphiphi}
\ee
Since both distributions are frequently mentioned in the literature, we hereafter for more certainty call $\varphi(m)$ the mass spectrum and $\phi(m)$ the mass function.

If one adopts a single completeness limit $\hat{d}_{xy,0}=1.8$ kpc, valid for clusters of all masses (horizontal line in Fig.~\ref{fig:dxy_lgm}), then $S(m) \equiv S_0=\pi\hat{d}_{xy,0}^2$, and Eq. \ref{eq:massdistr} re-written in the discrete form simply reflects the distribution of cluster numbers $\Delta_k N$ within the completeness area (i.e. below the horizontal line):
\be
\phi_k= \frac{1}{S_0}\,\frac{\Delta_k N}{\Delta_k \log m}\,,   \label{eq:mdc}
\ee
where the mass step ($\Delta_k\log m$) can be a variable.
 
In the case of the magnitude-dependent completeness limit, the cluster density will be computed as a sum of partial densities $\varsigma=1/(\pi\,\hat{d}_{xy}^2)$ of clusters located within their proper completeness limits given by Eq.~\ref{eq:dciksrel}, that is those with $d_{xy}\leqslant \hat{d}_{xy}$ (black dots in Fig.~\ref{fig:dxy_lgm}):
\be
\phi_k= \frac{1}{\Delta_k \log m}\sum_{i=1}^{\Delta_k N}\varsigma_i = \frac{1}{\pi\Delta_k \log m}\,\sum_{i=1}^{\Delta_k N}
\frac{1}{\hat{d}_{xy,i}^2}\,. \label{eq:lgmdv}
\ee
Here, we sum over the $\Delta_k N$ black dots within the mass interval of $\Delta_k \log m$. We note that in the case of the constant completeness limit $(\hat{d}_{xy,i}\equiv \hat{d}_{xy,0})$, Eq.~\ref{eq:lgmdv} is naturally reduced to Eq.~\ref{eq:mdc}.

\begin{figure}[t!]
   \centering
\includegraphics[width=0.82\hsize,clip=]{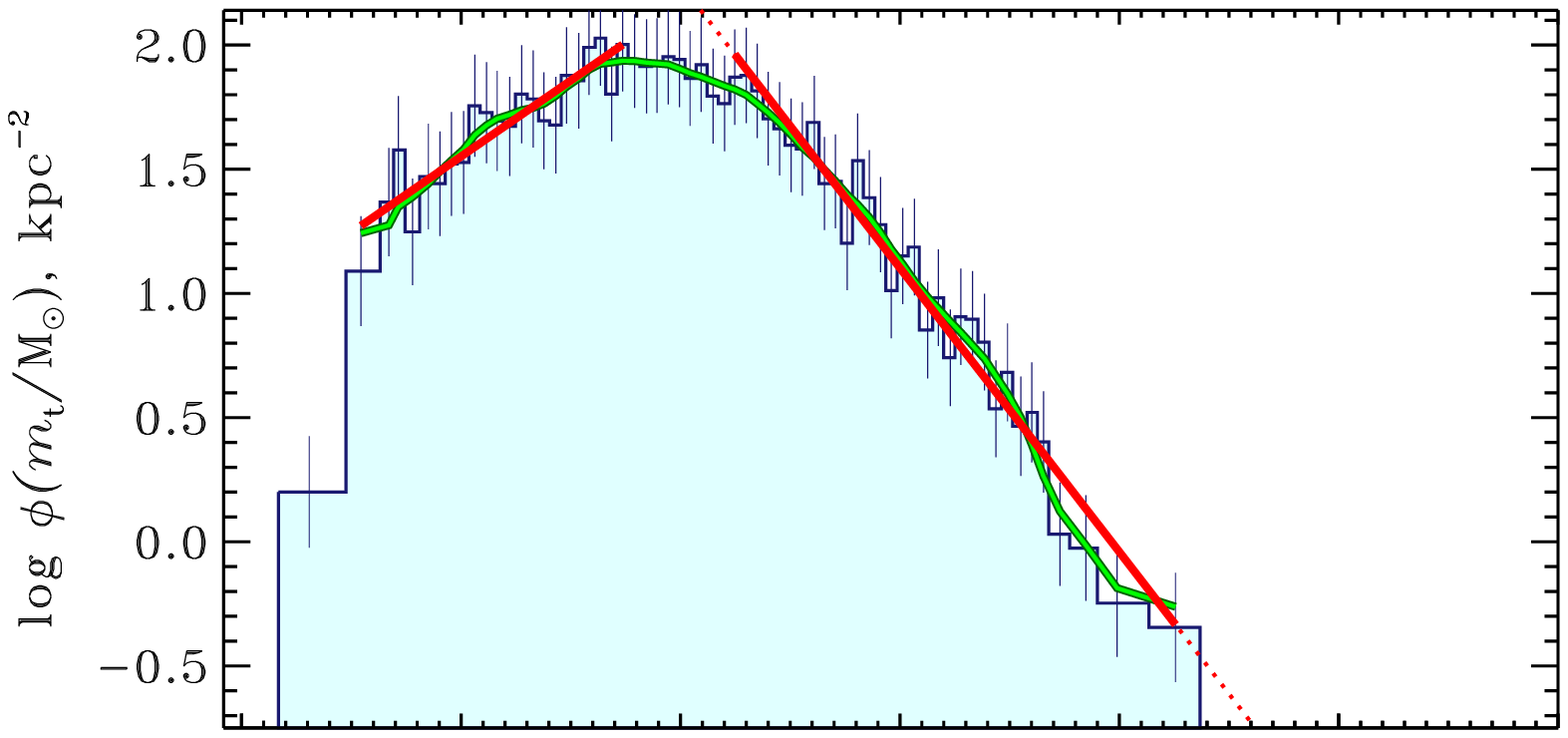}\\
\includegraphics[width=0.82\hsize,clip=]{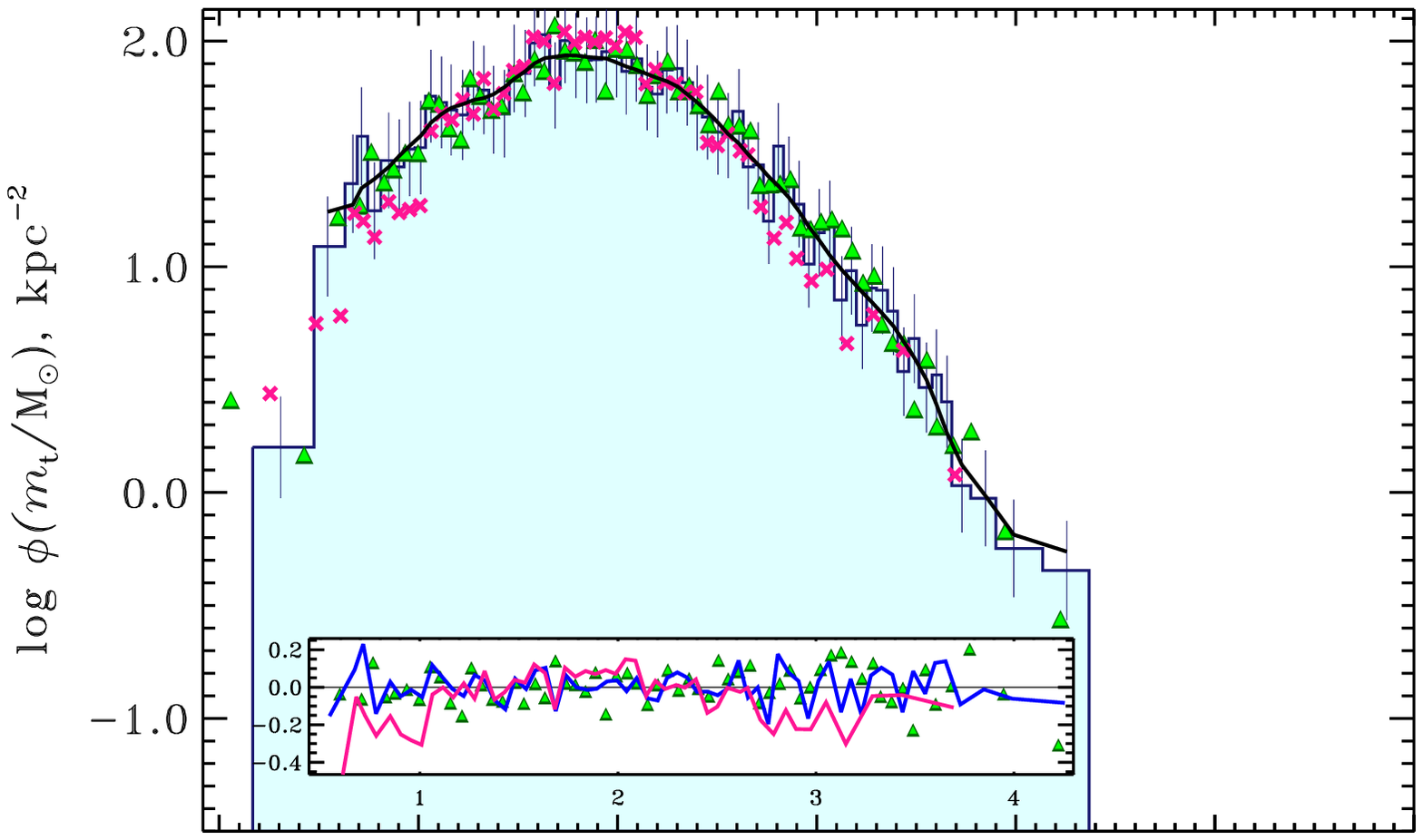}\\
\hspace{-0.75mm}
\includegraphics[width=0.825\hsize,clip=]{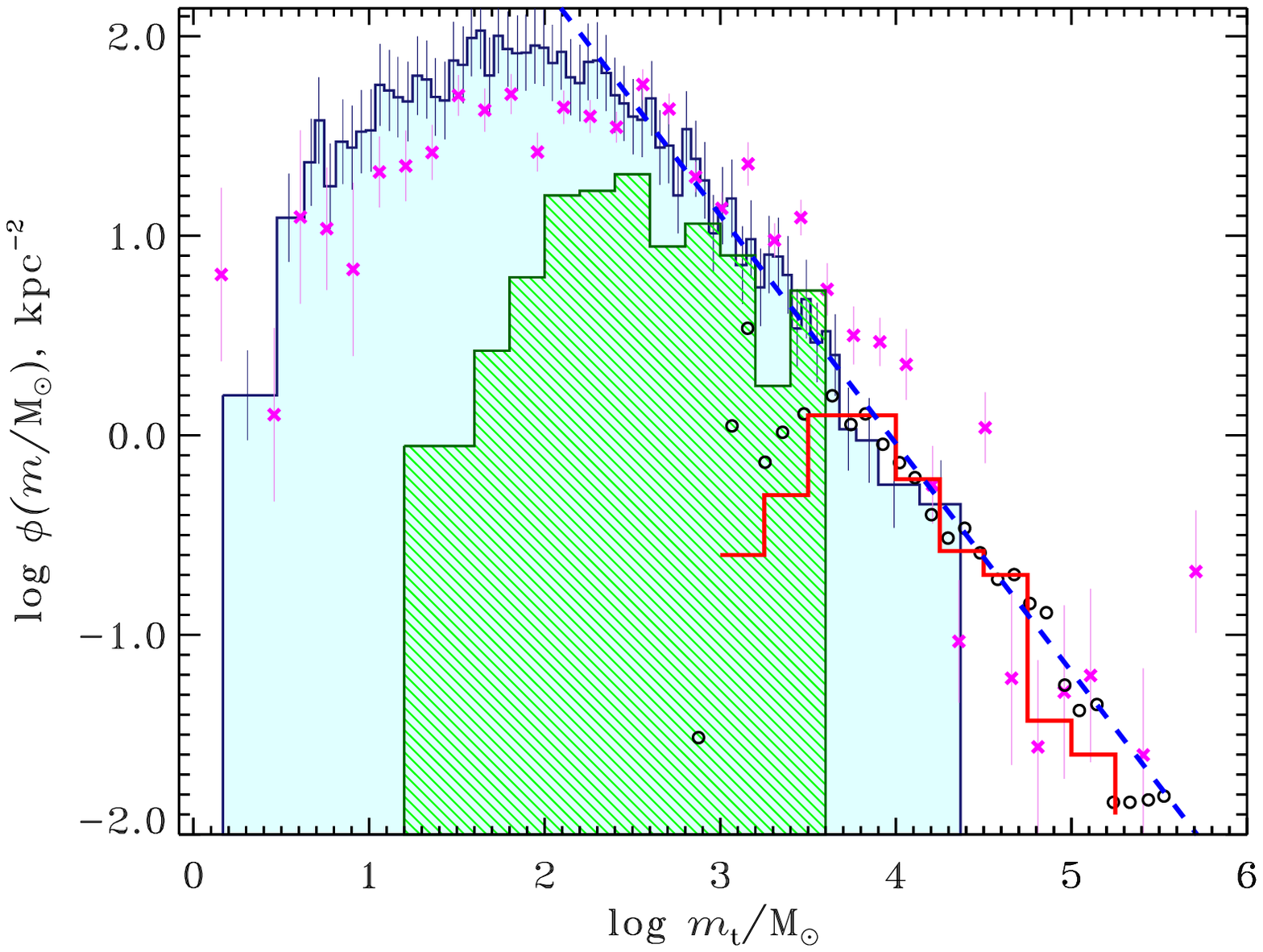}
\caption{GCMF computed with magnitude-dependent completeness limits in Eq.~\ref{eq:lgmdv} (light-blue filled histogram). Vertical bars show errors mainly due to uncertainty in mass. 
Top panel shows the low- and high-mass power-law fits (solid red lines) and the smoothed histogram data (green line). 
The middle panel presents different approaches to the mass function construction: single completeness limit with $\hat{d}_{xy,0}=1.8$ kpc (magenta crosses); local tidal mass (green triangles). The inset shows deviations from the smoothed data. 
Bottom panel presents a comparison with the literature CMFs: COCD-based for the Milky Way by \citet{lamea} (hatched histogram) and \citet{fuma} (crosses); for LMC star clusters by \citet{larss09} (open red histogram); for M83 by \citet{foue12} (open circles). The extragalactic CMFs are adjusted vertically for comparison. The dashed line shows the high-mass slope from the top panel.
}
\label{fig:pdfm}
\end{figure}

The resulting distribution, computed with the help of Eqs.~\ref{eq:mdc} and \ref{eq:lgmdv} with $\Delta_k N \geqslant 7$ and $\Delta_k \log m \geqslant 0.05$, is shown in Fig.~\ref{fig:pdfm}. In total, it consists of 2227 clusters residing within the completeness area, which is the wide solar neighbourhood shown with black dots in Fig.~\ref{fig:dxy_lgm}. Since it contains clusters of various ages, we call it the general CMF (GCMF).  The distribution extends over more than four decades in mass (between  $\log m_\mrm{t}/\mrm{M_\sun}=0.2$ and 4.4).  The mass distribution has a bell-like shape and resembles cluster mass functions observed in the Milky Way and in other galaxies (as discussed in more detail in Sect.~\ref{sec:cmplit}). The difference is in the position of the apparent maximum: in external galaxies, it depends on the distance to the galaxy, which implies that it is related rather to incompleteness beyond the observation limit. In contrast, in the GCMF built within the completeness zone, the position of the maximum manifests the details of cluster formation, evolution, and death.

The low-mass and high-mass branches of CMFs usually exhibit power-law shapes and characterised by the slope ($x$). We quantify them by linear fits as follows:
\be   
 \log \phi = \log \phi_0 - x\,\log m\,. 
 \label{eq:fitlaw}
\ee 
For the determination of fit parameters, we use the standard {\tt IDL} routine {\tt LINFIT}. Uncertainties in the mass of the clusters dominate errors in the input data, which were estimated using Monte Carlo simulations. We calculate the reduced goodness-of-fit statistic $\chi^2_0$ along with the $P$-value that gives the probability that the computed fit would have a value of $\chi^2_0$ or larger. If $P$ is greater than 0.1, the linear model parameters are `believable'. The mass intervals are selected to maximise the $P$-values.

The fit for the high-mass end is drawn in the range $\log m_\mrm{t}/\mrm{M_\sun}= 2.3-4.3$ (top panel of Fig.~\ref{fig:pdfm}). The parameters of the fit are zero point $\log\phi_0=4.53\pm0.21$ and slope $x= 1.14\pm0.07$. 
Substitution of Eq.~\ref{eq:varphiphi} for Eq.~\ref{eq:fitlaw} 
shows that the mass spectrum also follows a linear relation with a slope $\alpha=x+1= 2.14\pm0.07$. 
For the low-mass end, we find a rising slope with $x= -0.61\pm0.13$ (see also Table~\ref{tab:cmfitprm} in Sect.~6).

The tolerance of the mass function to some systematic effects is illustrated in the middle panel of Fig.~\ref{fig:pdfm}. It shows a function obtained with the assumption of a single completeness limit and a function built using tidal masses determined for parameters of the tidal field at the solar position (referred to as local tidal masses) for comparison. We can see that both effects have a small impact on the final mass function. The distribution of the local tidal masses perfectly fits the GCMF (most $\log m$-bins only differ within the statistical uncertainty due to binning effect). This can be explained by the fact that the dependence of the tidal parameters, $\Omega$ and $\beta,$ (cf.  Eq.~\ref{eq:jamas}) on the Galactocentric radius cancels out in the lowest order within the completeness area. Despite the considerable difference in the number of objects collected from different areas (1328 versus 2227), the disagreement between the GCMF and the single completeness limit mass function is on the order of the statistical uncertainty, although it seems to be systematic at the limits of the mass scale. We also note that the single completeness limit approach is unable to provide a reliable mass function at the high-mass end ($\log m_\mrm{t}/\mrm{M_\sun} >3.4$).

\subsection{Comparison with prior results}\label{sec:cmplit} 

In the Milky Way, the first attempt to build the luminosity function of the Galactic clusters from the literature data was undertaken by \citet{vdblaf84}.
It was based on a sample of 142 clusters that, according to the authors, is to two-thirds complete within 400 pc. 

A higher completeness degree within 650 pc was achieved later with our Hipparcos-based survey COCD \citep{clucat}. \citet{lamea} used a sub-sample of 114 COCD clusters within 600 pc from the Sun. For a cluster mass estimation, they integrated a \citet{salpeter55} stellar initial mass function (IMF) from the brightest end of the cluster Main Sequence down to an arbitrary common mass limit of $0.15\, \msun$. The IMF was properly normalised for every cluster based on the COCD data on the number of observed cluster $2\sigma$-members. They built the number distribution within 600 pc from the Sun and found that at the low mass end it is limited to 100 $\msun$.

The full COCD survey including the extension by \citet{newc109} was used to build cluster mass and luminosity functions by \citet{fuma}. The cluster masses were estimated from tidal radii determined from King profiles. The cluster distribution over apparent integrated magnitudes shows that the cluster sample is complete down to the apparent integrated magnitude $V=8$ mag, with 440 clusters above this completeness limit. This, on average, corresponds to a completeness area in the solar neighbourhood with an effective radius of about 1 kpc. The masses of the Galactic clusters span a range from a few solar masses to $\log m_\mrm{t}/\mrm{M_\sun} \approx 5.5$. 

Among extragalactic results, the following two could be used for our comparison purposes.
\citet{larss09}, in his study of the massive end of cluster mass functions in spiral galaxies, also built a mass function of star clusters in the Large Magellanic Cloud. He used literature data on \textit{UBV} photometry of 504 clusters and applied simple stellar population (SSP) models for the mass and age determination for clusters with ages $\log t/\mrm{yr}<9$.  The estimated completeness limit of the photometry is $V\approx 13$ mag, which corresponds to a limiting mass of $\log m/\mrm{M_\sun} \approx 4$. 

\citet{foue12} studied the age and mass distributions of 1242 star clusters in the central and north-eastern part of the galaxy M83, a nearby analogue of the Milky Way. The observed basis was provided by the Hubble Space Telescope (HST) aperture \textit{UBVIH}$_\alpha$ photometry corrected for foreground Galactic extinction. The cluster parameters were determined with the help of stochastic SSP models. The mass function was constructed for masses $\log m/\mrm{M_\sun}>3$. The mass spectrum at $\log m/\mrm{M_\sun}>3.5$ (which is close to the survey completeness limit at $V\approx 23$ mag) was fitted by a power law with a slope $\alpha=$ 2.15$\pm$0.14.

\begin{table*}[t!] 
 \caption{Parameters of cluster age groups. }
 \label{tab:agegprm}
\begin{center}
\tabcolsep=4pt
\begin{tabular}{rrcccc}
\hline\hline
\vgap
\mc{1}{c}{Age group} &     & Age interval & Average age & Age st. deviation & Mass range \\
 &  \mc{1}{c}{$N$}   &    $\Delta\log t/\mrm{yr}$    &$\langle\log t/\mrm{yr}\rangle$&$\sigma(\log t/\mrm{yr})$&$\Delta\log m_\mrm{t}/\mrm{M_\sun}$ \\
\vgap
\hline
\vgap
Initial-Age (I)&   234  &  6.4--7.3 & 6.95 &  0.26 &  0.5--4.4 \\   
Young (Y)&   503  &  7.3--8.3 & 7.90 &  0.29 &  0.5--3.8 \\   
Medium-Age (M)&  1051  &  8.3--9.1 & 8.75 &  0.22 &  0.2--4.3 \\   
Old (O)&   391  &  9.1--9.8 & 9.32 &  0.15 &  1.0--4.0 \\
\vgap
\hline
\vgap
General (G)&  2227  &  6.0--9.8 & 8.41 &  0.80 &  0.2--4.4 \\
\vgap
\hline
\end{tabular}
\end{center}
\end{table*}

The comparison of the above distributions with the GCMF is shown in Fig.\ref{fig:pdfm} (bottom panel). To convert the \citet{lamea} counts into surface densities, we divided their data by the area of the circle with a radius of 0.6 kpc. \citet{lamea} data fit the GCMF at $\log m_\mrm{t}/\mrm{M_\sun}=$ 3.0--3.5. At lower masses, their distribution shows increasing deficiency compared to our GCMF due to incompleteness. At higher masses there is a cut-off bias in the \citet{lamea} sample due to low-number statistics.
For the full COCD sample, the derived CMF agrees reasonably well with the GCMF over the total range of masses \citep{fuma}. The observed scatter is significantly larger due to the smaller sample size.
The counts of \citet{larss09} and \citet{foue12} were adjusted vertically to be comparable to the GCMF. One can see that the slopes of both samples at $\log m_\mrm{t}/\mrm{M_\sun} > 3.5$ are consistent with the GCMF.
Our general conclusion is that the agreement between the GCMF and the literature on the MW and some nearby galaxies is satisfactory.

\section{Temporal and spatial variations of the CMF}\label{sec:tvar}

The GCMF  that we describe is a product of several processes: the formation of clusters, their dynamic evolution associated with mass loss and disintegration, as well as the evolution of the Galactic disc as a whole by, for instance, changes in the rate of star formation with time. As a result, we observe as the GCMF a certain integral including all these processes. In this section, we study the age dependence of the CMF over the entire period of cluster existence.

\begin{figure}[b!]
   \centering
\includegraphics[width=0.9\hsize,clip=]{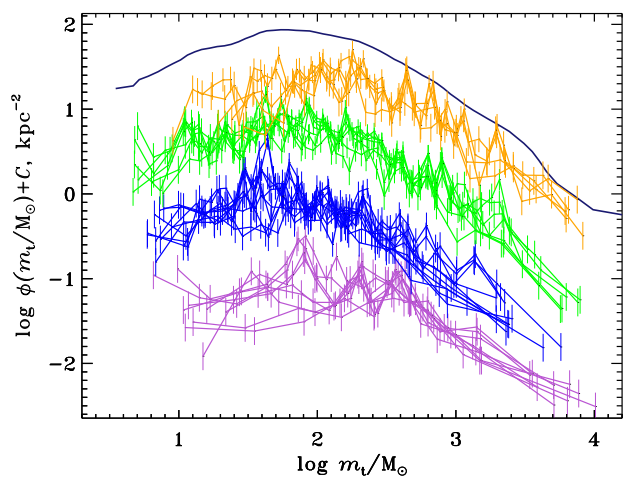}
\caption{Comparison of selected CMF age groups
built as described in the text. The groups are marked with different colours and are artificially shifted vertically to reduce the overlap. From bottom up, we show the initial-age (violet), young (blue), medium-age (green), and old (orange) cluster groups. The uppermost dark curve is the GCMF (including all ages) constructed in Sect.~\ref{sec:method}. The groups and their age limits are presented in Table~\ref{tab:agegprm}.
}
\label{fig:mf_groups}
\end{figure}

\subsection{Age group definition}\label{sec:grpdef}

Considering the mass functions of clusters of different age groups, we noticed that although they may share a  resemblance with the GCMF, they nevertheless slowly change their morphology with age, an attribute that appears to be associated with the evolution of clusters. Since these variations are insignificant and it is unclear  a priori  how they may be formalised, we first scanned a complete sample of clusters over the entire interval of available ages, $\log t/\mrm{yr}=$ 6.4--9.6, using the sliding window method with a carriage width of $\Delta\log t/\mrm{yr}=0.4$ and a step of $\delta\log t/\mrm{yr}=0.1$.   For every sub-sample falling into the window, we built the respective mass function. We limit ourselves by this lower limit of the ages since: i) it is about the original young limit of Padova isochrones available for $\log t/\mrm{yr} \geqslant 6.6$; and ii) we isolate in this way the embedded phase of cluster evolution assumed to last for 3 Myr \citep{parmen13}.  As the analysis has shown, the constructed family of mass functions can be divided into several groups based on the CMF features in the region of moderate and small masses. It turns out that this classification correlates well with the age of the clusters. In Fig.~\ref{fig:mf_groups}, we show the obtained mass function groups. Curves belonging to our classification to the same age group have the same colour.

We attribute the youngest group (designated with I, which stands for 'initial-age' clusters) to clusters with a long and almost flat low-mass segment ($\log m_\mrm{t}/\mrm{M_\sun} \lesssim 2.7$). In the other age groups, this section is no longer flat, but has a noticeable maximum, with the position being shifted with age toward larger masses. According to the position of the maximum,
we distinguish a young group (Y, representing 'young' clusters) with a maximum at $\log m_\mrm{t,max}/\mrm{M_\sun} \simeq 1.7$, a medium group (M, for 'medium-age' clusters) with $\log m_\mrm{t,max}/\mrm{M_\sun}  \simeq 1.8$, and an old group (O,
for 'old' clusters) with $\log m_\mrm{t,max}/\mrm{M_\sun} \simeq 2.1$. The dependence of the position of the CMF maximum on time leads to the appearance of a broad maximum at the integral function, which is the GCMF. Table~\ref{tab:agegprm} gives the number of clusters in each group, the age limits including mean age and standard deviation, and the mass range covered. The youngest 48 clusters with $\log t/\mrm{yr}<$ 6.4 are excluded from the age groups.

\subsection{CMFs at different age groups}\label{sec:cmft}

\begin{figure*}[t!]
   \centering
\includegraphics[width=0.525\hsize,clip=]{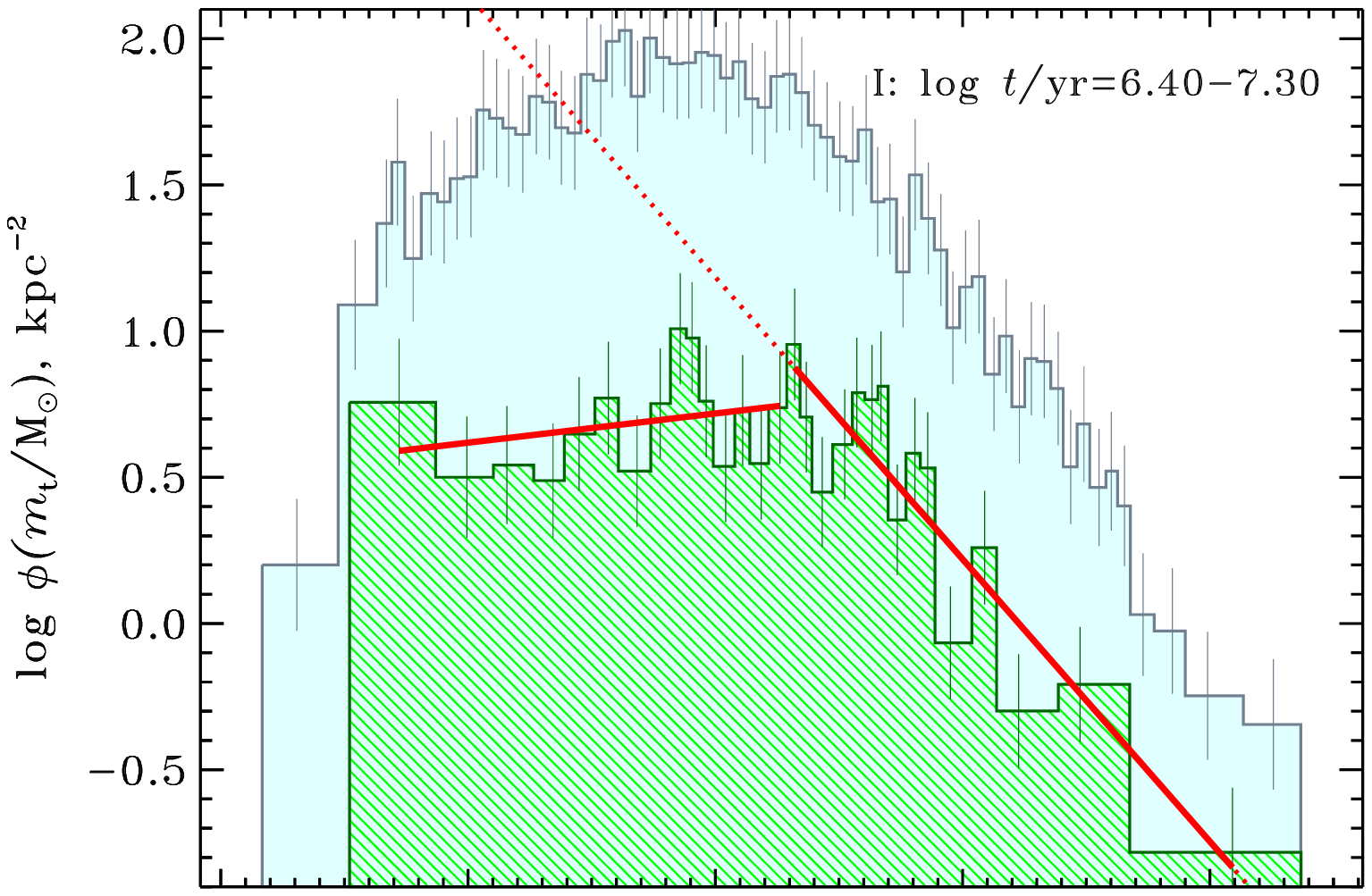}
\includegraphics[width=0.45\hsize,clip=]{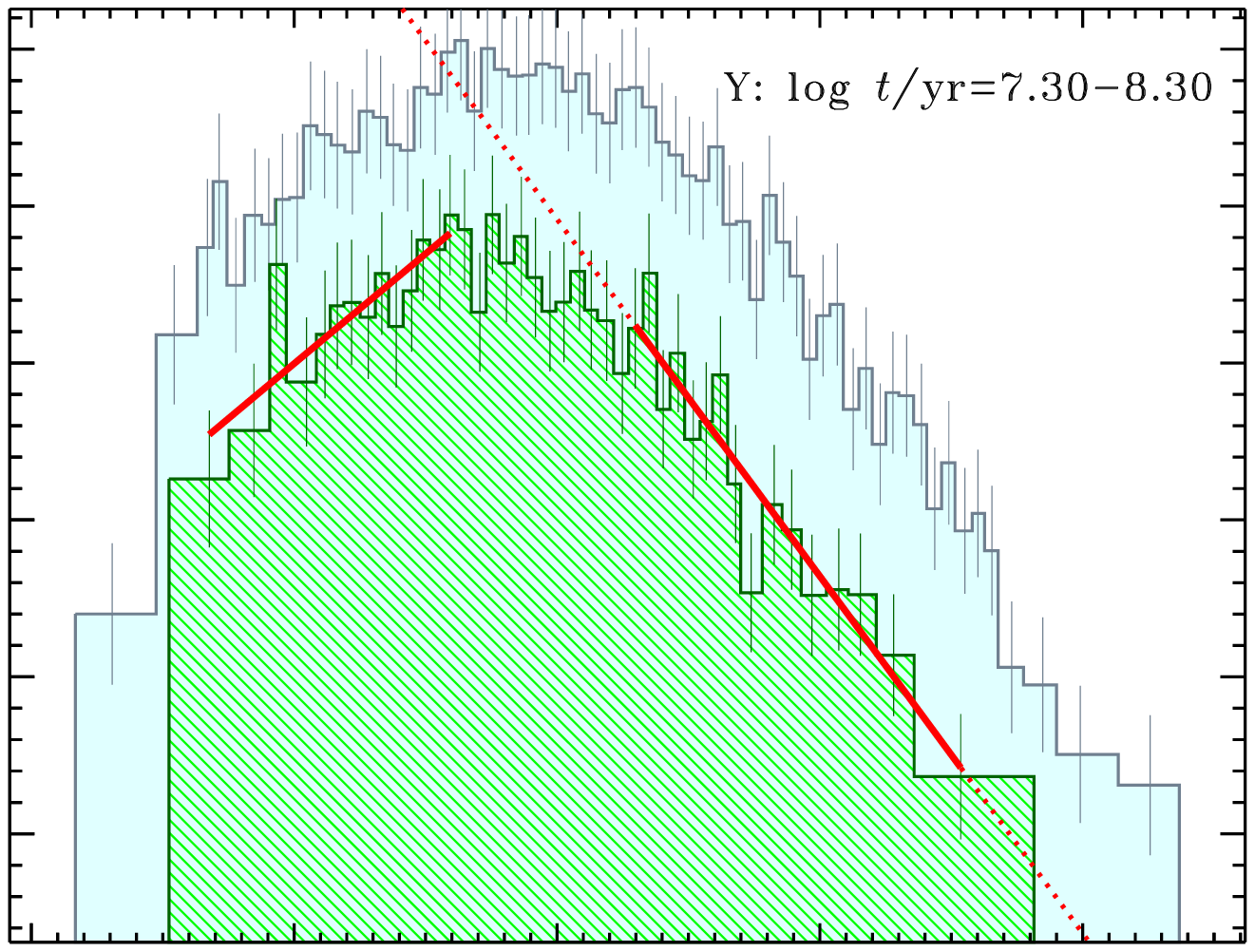}
\includegraphics[width=0.525\hsize,clip=]{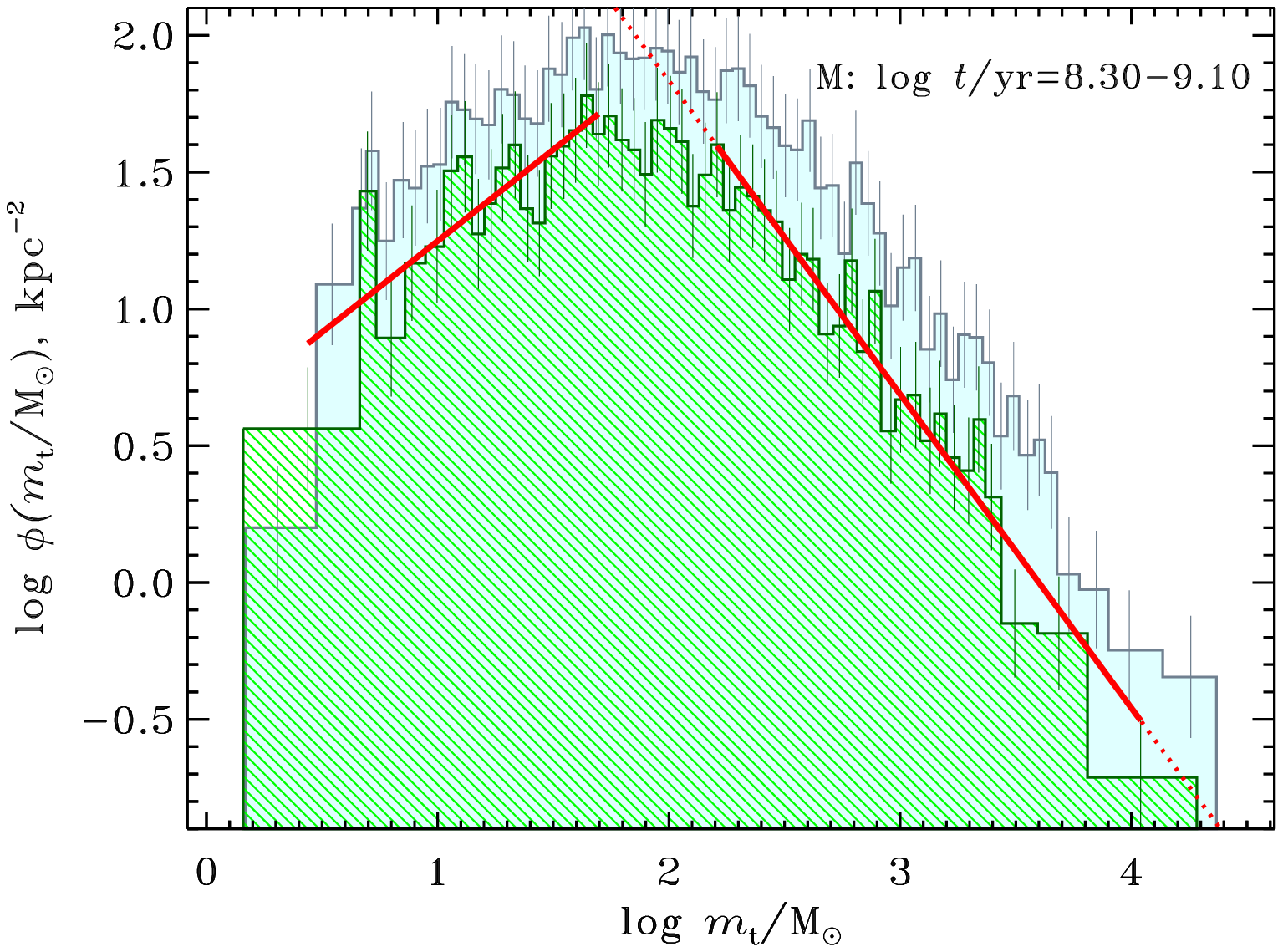}
\includegraphics[width=0.45\hsize,clip=]{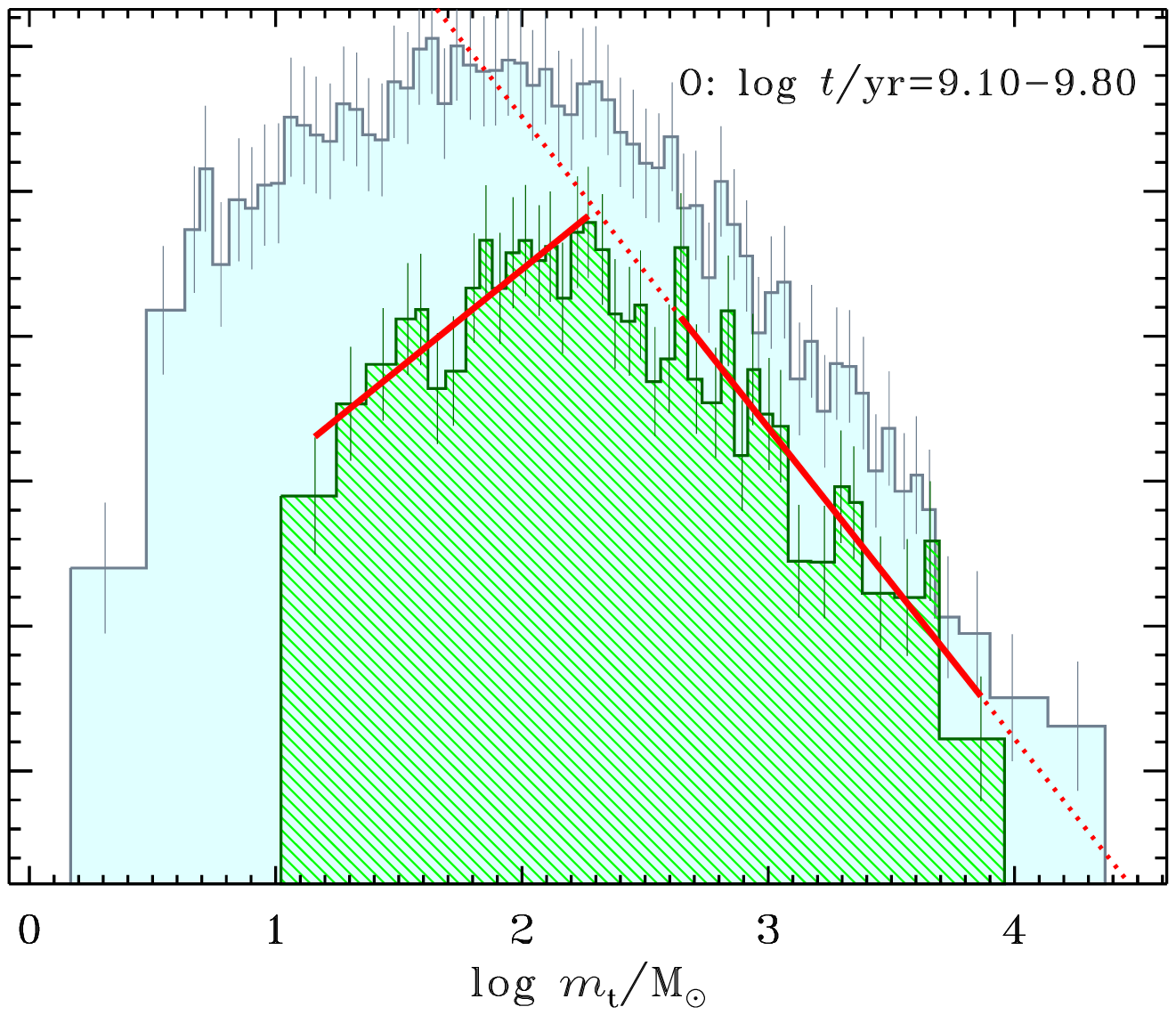}
\caption{Mass functions of star clusters of four age groups (group identification and age limits are shown in the panels). The CMFs are given by green histograms, while the background shows the GCMF. The straight lines represent the linear fit of the CMFs with power laws. The mass intervals used for the fit determinations are shown by solid red lines. 
}
\label{fig:mf_iy}
\end{figure*}

The CMFs for different age groups are built in the same way as the GCMF described in Sect.~\ref{sec:clumf}. The resulting CMFs (shown in Fig.\,\ref{fig:mf_iy}) can be characterised by power laws at the low- and high-mass ends. These segments are connected by a weakly pronounced maximum with varying positions and widths in the different age groups. The maxima of the distributions tend to have higher masses with age in groups Y, M, and O. The power-law fits were done as for the GCMF (Sect.~\ref{sec:method}).
Table~\ref{tab:cmfitprm} contains the mass intervals in which the fits are performed (Col.~3), the derived fit parameters and their errors (Cols.~4 and 5), and the fit quality parameters, $\chi^2_0$ and $P$. The last two lines related to the GCMF (Sect.~\ref{sec:method}) are given for comparison. 

\begin{table*}[tbh] 
 \caption{Power-law fit parameters (Eq.~\ref{eq:fitlaw}) of the cluster mass function for different age group segments.}
 \label{tab:cmfitprm}
\begin{center}
\tabcolsep=6pt
\begin{tabular}{cccrrccc}
\vgap
\hline\hline
\vgap
Age group & Segment & Fit interval &  \mc{2}{c}{Fit parameters} & \mc{2}{c}{Fit quality}  \\
&&$\Delta\log m/\msun$&\mc{1}{c}{$x$}&\mc{1}{c}{$\log\phi_0$}& $\chi^2_0$ & $P$ \\
\hline
\vgap
\multirow{2}{*}{Initial-age} &I$_0$& 0.7--2.3 &$-0.10\pm$0.12& 0.52$\pm$0.21 &0.79  &  0.74  \\
&I$_1$& 2.3--4.1 & 0.97$\pm$0.12& 3.12$\pm$0.33 &1.29  &  0.28   \\[1mm]
\multirow{2}{*}{Young} &Y$_0$& 0.7--1.6 &$-0.70\pm$0.21& 0.30$\pm$0.26 &0.47  &  0.95 \\
&Y$_1$& 2.3--3.6 & 1.14$\pm$0.14& 3.73$\pm$0.39 &0.64  &  0.87 \\[1mm]
\multirow{2}{*}{Medium-age}&M$_0$& 0.4--1.7 &$-0.67\pm$0.14& 0.58$\pm$0.18 &0.78  &  0.76  \\
&M$_1$& 2.2--4.1 & 1.15$\pm$0.08& 4.13$\pm$0.25 &0.55  &  0.97 \\[1mm]
\multirow{2}{*}{Old} &O$_0$& 1.2--2.3 &$-0.69\pm$0.15&$-0.14\pm$0.27 &0.48  &  0.96  \\
&O$_1$& 2.7--3.9 & 1.07$\pm$0.14& 3.90$\pm$0.45 &1.07  &  0.46 \\[1mm]
\hline
\vgap
\multirow{2}{*}{General} &G$_0$& 0.5--1.7 & $-0.61\pm$0.13& 0.94$\pm$0.16 &0.18  &  1.00  \\
&G$_1$& 2.3--4.3 & 1.14$\pm$0.07& 4.53$\pm$0.21 &0.40  & 1.00  \\
\vgap
\hline
\vgap
\end{tabular}
\end{center}
\end{table*}

The slopes of the high-mass sections (subscript 1) are found near $x = 1.15$ and they are almost independent of age, except for the youngest group (I), for which the slope is slightly shallower. The slope of the low-mass end also does not depend on age for groups Y, M, and O and is close to $x = -0.7$. For group I, the CMF is remarkably flat up to $\log m_\mrm{t}/\mrm{M_\sun}\approx 2.3$.

In our earlier work based on COCD \citep{fuma} it was found that the CMF for the youngest clusters ($\log t/\mrm{yr}\leqslant 6.9$) takes the form of a two-section distribution with a quasi-flat slope of $x=$ $-$0.18$\pm$0.14 at the low-mass end ($\log m_\mrm{t}/\mrm{M_\sun}=$ 1.7--3.4) and with a slope of $x=$ 0.66$\pm$0.14 at the high-mass end ($\log m_\mrm{t}/\mrm{M_\sun}=$ 3.4$-$4.9). The latter steepens with age up to $x=1.17$.  

\cite{bikea03} constructed mass functions from several hundred clusters observed in the region of the inner arms of the galaxy M51. According to the data of broadband photometry \textit{UBVRIH}$_\alpha$ and narrow-band indices OIII over an area of $\sim3\times3$~kpc, they constructed a mass distribution of 354 objects younger than $\log t/\mrm{yr} = 7.0$ in the mass range $\log m/\mrm{M_\sun}=$ 2.3--5. The masses are determined from the data on SSP models for \textit{UBVR} photometry with completeness for $\log m/\mrm{M_\sun}\gtrsim 3$. The resulting slope of the mass spectrum for 149 objects from this mass interval is $\alpha = 2.16$, which corresponds to $x=1.16$.

\cite{dowell08} constructed the mass spectrum of young clusters $(\log t/\mrm{yr}<7.3)$ from the SDSS survey in 13 nearby irregulars and 3 spiral galaxies. Cluster parameters (including absolute magnitudes, interstellar extinction, age, and mass) were determined from the colour charts of the $ugriz$ system. A total of 321 and 358 clusters were used in irregular and spiral galaxies. The mass range covers $\log m/\mrm{M_\sun}\sim$ 4.2--6.6 and the slopes ($\alpha$) are equal to 1.88$\pm$0.09 and 1.75$\pm$0.06, respectively. 

\cite{fallea05,fallea09} implemented a simple model of the formation and evolution of star cluster population to the observations of the HST in the Antenna galaxies (NGC~4038 and NGC~4039). Observations cover the main body of both galaxies. Cluster parameters 
such as age or absorption were determined from \textit{UBVIH}$_\alpha$ photometry, using \cite{bruz03} SSP models with Salpeter IMF and mass from $m/L_{\rm V}$ ratio. To reduce star contamination with bright stars the sample was cut (objects fainter than $L_{\rm V}/L_\sun< 3\times10^5$ are removed), which
limits cluster masses to $\log m/\mrm{M_\sun}>5.3$, with about 2300 clusters. The mass and luminosity functions were constructed for age intervals $\log t/\mrm{yr}=$ 6--7 and 7--8. It was found that for the mass interval $\log m/\mrm{M_\sun} =$ 4.5--7, the slope is $\alpha=$ 2.14$\pm$0.03 and 2.03$\pm$0.07, respectively.

\cite{chan10} studied the two-dimensional (2D) mass and age function of star clusters in the Large and Small Magellanic Clouds. They constructed mass distributions of clusters in different age ranges. To do this, they used \cite{huntea03} integrated \textit{UBVR} photometry (854 clusters in an area of 11 kpc$^2$ in the LMC and 239 clusters in an area of 8.3 kpc$^2$ in the SMC). Masses are determined using the $m/L_{\rm V}$ ratio and ages are determined from SSPs of \cite{bruz03} with a Salpeter IMF. Due to the proximity of the Magellanic Clouds, the work extends the range of masses and ages of clusters compared to those from the Antennae galaxies previously studied by the authors. It was found that the obtained distributions are close to similar distributions in the aforementioned galaxies, representing another more massive class of stellar systems. 
As \cite{chan10} established for $\log t/\mrm{yr}\leqslant 9$, the slope ($\alpha$) of the constructed mass functions is practically independent of age and is equal to 1.8$\pm$0.2 in both LMC and SMC. 

\cite{larss09} constructed the CMF of star clusters in several spiral galaxies and considered the possibility of representing them as a distribution within the framework of the Schechter approximation~\citep{schechter76}. Using selected spiral galaxies and SSP models, the CMF was constructed for young ($\log t/\mrm{yr}<8.3$) clusters in rich (NGC~5236, NGC~6946 having more than 100 objects per galaxy) and poor (more than 40 objects per galaxy, cf. Table~1 therein) spirals. Observations span over $\log m/\mrm{M_\sun}=$ 4.0--6.0, the masses are determined with the help of SSP models and photometric data, the count completeness is expected at $\log m/\mrm{M_\sun}>5$. 
For these masses, the data are compatible with a slope $\alpha=2.0$ and Schechter cut-off of $M_c=2.1\times 10^5\,\mrm{M_\sun}$.
The comparison given above shows that the new data supports our previous findings on the CMFs slopes for different age groups in the Milky Way and are similar to the high-mass end slopes found for extragalactic systems.

\begin{figure}[b!]
   \centering
 \includegraphics[width=0.99\hsize,clip=]{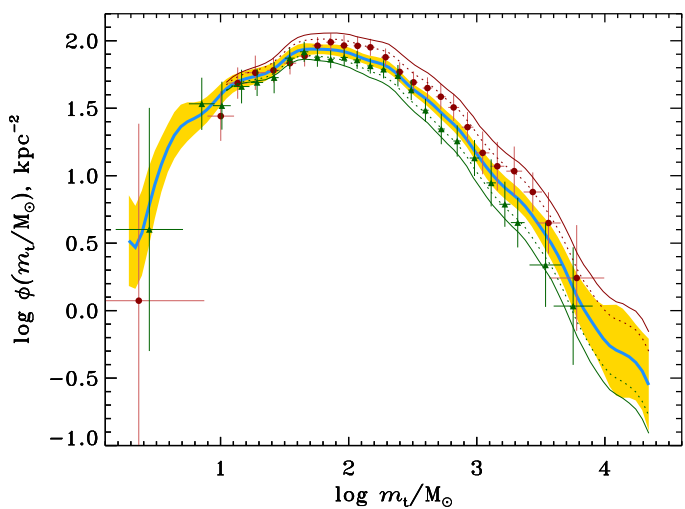}
 \caption{CMFs of star clusters for two ranges
of Galactocentric radii, as presented in Table\,\ref{tab:tabpq}: red dots are for the inner sub-sample and green triangles are for the outer sub-sample. The vertical bars are statistical errors characterising a bin population and the horizontal bars indicate the bin widths. The bold blue line is the smoothed GCMF derived in Sect.~\ref{sec:method} and the yellow background highlight represents its statistical errors. Thin red and green solid and dotted lines represent GCMFs of the inner and outer sub-samples after applying the bias due to the exponential decline of the Galactic disc surface density with a scale length of 3.8 and 6~kpc, respectively.  }
 \label{fig:spatial}
 \end{figure}

\subsection{CMFs at different Galactocentric radii}\label{sec:cmfr}

In the sections above, we assume that there is independence 
of the completeness limit with regard to the direction of observations and,
thus, the uniformity of the clusters' distribution in the Galactic plane as well. In this sub-section, we address this assumption in more detail. 

Since CMFs depend on the disc surface density, we would expect larger and lower values for the inner and outer sub-samples, respectively, compared to the GCMF. The difference should be more pronounced at the high-mass end resulting in a bias in the shape of the CMFs because massive clusters are seen reliably at larger distances. In the following, we evaluate the impact of a radially decreasing cluster surface density on the CMF 
by selecting an inner ($R_{\rm \odot}-R>$ 0.35 kpc) and outer ($R-R_{\rm \odot}>$ 0.35 kpc) sub-samples of clusters.
We excluded the region $|R-R_{\rm \odot}|\le 0.35$ kpc hosting 450 clusters to increase the contrast between the inner and the outer sub-samples. 

\begin{table*}[t!]
\tabcolsep=6pt
 \begin{center}
 \caption{Updated parameters of the `completeness distance-magnitude' relation (Eq. \ref{eq:dciksrel}).}\label{tab:tabpq}
\begin{tabular}[c]{cccccr}
\hline
\hline
\vgap
Sample& $p$,   & $q$,                  & Range, & Median position,     & $N_{\rm obj}$ \\
      & kpc   & kpc$\cdot$mag$^{-1}$ & kpc    & kpc &           \\
\vgap
\hline
\vgap
Inner sub-sample      & 0.36  & $0.48$ & 4.2--8.1  &  7.4  & 774 \\
Outer sub-sample      & $-0.05$ & $0.69$ & 8.9--13.5 &  9.8 & 1028 \\ 
\vgap
\hline
\vgap
General     & 0.36  & $0.54$ & 3.9--13.2 &  8.6  & 2227\\
\vgap
\hline 
\end{tabular}
 \end{center}
\end{table*}

\citetalias{mwscint} has shown that the parameters $(p,q)$ for the completeness limit (Eq. \ref{eq:dciksrel}) depend on the direction in the Galaxy. Here, we determine these parameters separately for the inner and outer sub-samples. Following the procedure given in Sect. 4.2 of Paper~V, we recalculated the parameters $(p,q)$ given therein by applying additional smoothing with the sliding window method (as done for the total set of clusters in Sect.~\ref{sec:clumf}). The new parameters and the radial ranges for the general,  inner, and outer sub-samples are given in Table~\ref{tab:tabpq}. They describe a somewhat stronger dependence of the completeness limit on the integrated magnitude $I(M_{{K}_{\rm S}})$, compared to the one obtained in Paper~V. 
Using the newly derived parameters $(p,q)$, we redefined the completeness subsets for the inner and outer sub-samples. 
The number of clusters ($N_{\rm obj}$) in the respective magnitude-limited completeness ranges are given in Table~\ref{tab:tabpq}.  We note that the inner and outer sub-samples do not coincide with the clusters in the corresponding Galactocentric ranges of the general sample due to the different completeness limits.
A comparison of the respective CMFs (red circles and green triangles) with the GCMF constructed in Sect.~\ref{sec:method} is presented in Fig.~\ref{fig:spatial}. As expected, the vertical offsets are more pronounced at the high-mass end.

To calculate the impact of a Galactic cluster surface density profile, we adopted a universal shape of the GCMF and applied an average individual radial offset of each mass bin. Solid red and green lines in Fig.~\ref{fig:spatial} represent the resulting CMFs of the inner and outer sub-samples, respectively, assuming a radial scale length equal to 3.8~kpc \citep[as found for the young disc population by][]{2017A&A...602A..67A}. The dotted red and green lines are calculated for a radial scale length of 6~kpc.
We can see that the observed CMFs of star clusters of the inner and outer sub-samples (red circles and green triangles) agree well with a universal shape of the GCMF distorted by the radial gradient of the cluster surface density. 

In summary, the full cluster sample with magnitude-dependent completeness limits and the corresponding GCMF is representative of the open cluster population in the wider solar neighbourhood, despite the statistical impact of a radial decrease of the Galactic cluster surface density.

\section{Cluster formation and evolution model}\label{sec:model}

In Sect.~\ref{sec:cmft}, we discuss and quantify the CMFs of different age groups (see also Fig.~\ref{fig:mf_iy}). 
In this section, we present a simple model of cluster formation and its subsequent evolution that reasonably reproduces the obtained mass functions. At this stage, parameters for the model were found by hand, without using any best-fitting algorithm.

For CMFs at different ages, we need the CIMF, the cluster formation rate (CFR), and the cluster bound-mass function $m(M,t),$ depending on the initial mass ($M$) and age ($t$). In the subsections below, we derive the surface density of clusters from the data and describe our model in detail. All parameters adopted for the model are compiled in Table \ref{tab:model}.

\subsection{Surface density of clusters}\label{sec:cmf}

The number surface density of clusters as a function of current mass ($m$) and age ($t$) is given by:
\be \label{eq:sigma}
\sigma(m,t) \,\d m \d t = \Psi(t)f(M)\, \d M \d t\,.
\ee
Here, $\Psi(t)$ is the CFR as a function of age and $f(M)$ is the CIMF. To convert the initial mass bins to bins of the current mass, we need to know the bound-mass function $m(M,t)$. From Eq.~\ref{eq:sigma}, we obtain:
\be \label{eq:sigma2}
\sigma(m,t) = \Psi(t) \, f(M) \, \left(\frac{\p m(M,t)}{\p M}\right)^{-1} \,.
\ee
Here, it is assumed that the partial derivative is positive and does not vanish (see also discussion in Sect.~7.4). Then one can invert the bound-mass function in favour of the initial mass $M=M(m,t)$ for fixed age ($t$). However, if $m(M,t)$ is a non-monotonic function of $M$, then there is no one-to-one correspondence between initial and current masses. Some values of the current mass ($m$) can be obtained for several initial masses, $M_k(m,t)$. Equation~\ref{eq:sigma2} can then be modified as follows:
\be \label{eq:sigma2m}
\sigma(m,t) = \Psi(t) \, \sum\limits_k f(M_k) \, \left|\frac{\p M_k(m,t)}{\p m}\right|\,.
\ee
The cluster mass spectrum $\varphi(m)$ 
is obtained for a given age range $t_0<t<t_1$ by integrating the cluster surface density over age:
\be \label{eq:varphi_model}
\varphi(m) = \int_{t_0}^{t_1} \sigma(m,t)\, \d t\,.
\ee
Lastly, the total number surface density for this age range is given by:
\be \label{eq:Sigma_model}
\Sigma_{\rm N} = \int \varphi(m)\, \d m\,.
\ee

\subsection{Cluster formation rate}\label{sec:cfr}

We are interested in the number surface density of clusters in the solar neighbourhood. Therefore, the unit of the cluster formation rate $ \Psi(t)$ in the Galactic plane is the number of clusters per square kpc and Myr. Since the CFR is strongly degenerate with the cluster mass loss function, it is sufficient for our simple model to use a constant CFR, given by:
\be \label{eq:CFR}
\Psi(t) = \beta\,.
\ee
The value of $\beta$ is adapted to reproduce the total number density of observed clusters in the mass range of $\log m/\msun = 0-4.5$ and the age range of $\log t/\mrm{yr} = 6.5-10$.

\subsection{Cluster initial mass function}\label{sec:cimf}

In principle, the CIMF should be consistent with the observed present-day cluster mass function for the youngest clusters. The observations, however, may be strongly biased for several reasons.
Clusters up to an age of a few Myr could still be embedded in their parent molecular cloud and may even still form stars. Furthermore, infant mortality and the very fast evolution in the violent relaxation phase (see next subsection) influence the cluster mass function heavily.

Nevertheless, the two power-law segments shown in the top left panel of Fig.~\ref{fig:mf_iy} for the initial-age group suggest that we may use a  broken-power-law CIMF. Thus, we assume a two-slope broken power law with a smooth transition and an additional exponential Schechter-like cut-off for the largest masses required by the lack of clusters with masses above 50\,000$\,\msun$:
\be
f(M) = \frac{\d N}{\d M} = k_0 
\left( \displaystyle\frac{M}{M_\star} \right)^{-(x_1+1)}
\left[ 1+\left(\displaystyle\frac{M}{M_\star}\right)^s 
\right]^{\textstyle\frac{x_1-x_2}{s}}
  \exp\left(-\frac{M}{m_{\rm S}}\right) \,.
\label{eq-N}
\ee
Here, $x_1$ and $x_2$ are low- and high-mass power-law indices, $M_\star$ and $s$ determine the position and sharpness of the transition, and $m_{\rm S}$ is the characteristic mass for the Schechter cut-off. The normalisation constant, $k_0$, 
normalises the CIMF, $f(M)$, to unity
for the chosen lower cluster mass limit $m_\mrm{lower} = 2\msun$. 

\begin{table}[t!]
\tabcolsep=6pt
\caption {Parameters of cluster formation and evolution model. The meaning of each parameter is described in the text.}\label{tab:model}
\centerline{
\begin{tabular}{c c}
\hline 
\hline
\vgap
Parameter & Value\\
\vgap
\hline
\vgap
\multicolumn{2}{c}{{\bf - CFR  (Eq. \ref{eq:CFR})}}\\
\vgap
\vgap
$\beta$ & 0.81 kpc$^{-2}$ Myr$^{-1}$ \\ 
\vgap
\hline
\vgap
\multicolumn{2}{c}{{\bf - CIMF (Eq. \ref{eq-N})}}\\
\vgap
$k_0$ & $1.5\times 10^{-4}\,\msun^{-1}$ \\ 
$M_\star$ & 1000 $\msun$\\ 
$s$ & 2.4 \\ 
$x_1$ & 0 \\ 
$x_2$ & 1.2 \\
$m_\mrm{S}$ & 85000 $\msun$ \\ 
\vgap
\hline
\vgap
\multicolumn{2}{c}{{\bf  - stellar evol. $\mu(t)$ (Eq. \ref{eq:mu})}}\\
$p_0$ & 1.0078 \\
$p_1$ & $-0.07456$ \\
$p_2$ & $-0.02002$ \\
$p_3$ & 0.00340 \\
\vgap
\hline
\vgap
\multicolumn{2}{c}{{\bf - violent relax. (Eq. \ref{eq:nv})}}\\
$n_\mrm{b}$ & 0.1 \\
$t_\mrm{v}$ & 5 Myr \\
\vgap
\hline
\vgap
\multicolumn{2}{c}{{\bf- cluster dissolution  (Eqs. \ref{eq:dnd} to \ref{eq:lt})}}\\
$a_1$ & $-0.2$ \\
$a_2$ & 0.9 \\
$c$ & 0.23 Gyr$^{-1}$ \\
$M_\mrm{br}$ & 5000 $\msun$\\
\vgap
\hline
\end{tabular}}
\end{table}

\subsection{Cluster bound-mass function}\label{sec:mt}

Three processes affect the bound mass of the cluster: the mass loss during violent relaxation, the relatively slow destruction of the cluster in the Galactic tidal field, and the stellar evolution of stars within the cluster. The first two effects reduce the number of stars, whereas the third reduces the masses of stars, not their number. Stellar evolution depends on the initial mass function within the cluster and the metallicity. According to \citet[][Table~B2]{lamea10}, the normalised stellar mass $\mu(t)$ as a function of age can be approximated by the third-order polynomial 
\be
\mu(t) = \sum_{i=0}^3 p_i\,\Bigr[\log (t/\mrm{Myr})\Bigr]^i.
\label{eq:mu}
\ee
In our model, we assume a metallicity of $Z=0.008$ for the coefficients $p_i$ (see Table~\ref{tab:model}).
The time scales of the dynamical processes mentioned above are clearly separated. Here, we used the N-body simulations by \citet{shukirea17,shukirea18} as an orientation. The initial conditions of these models were based on a centrally peaked star formation efficiency by adopting 5\% star formation efficiency per free-fall time. Starting with a Plummer model \citep{plummer11} for the star cluster, the gas profile before instantaneous gas removal was determined and the cluster was initialised in dynamical equilibrium including the gas potential. Instantaneous gas expulsion leads to a supervirial cluster, which quickly expands and loses a large fraction of stars. This phase is called `violent relaxation' and takes 10--20~Myr. Above a threshold of 13\% for the global star formation efficiency, a core of stars re-collapses and forms a bound cluster in dynamical equilibrium. The fraction of stars that remain in the cluster depends strongly on the global star formation efficiency, but not on the initial mass and weakly on the Roche volume filling factor. In our model, the decrease in the number star fraction due to violent relaxation is governed by the function:
\be
n_\mrm{v}(t) = n_\mrm{b} + (1-n_\mrm{b})\cosh^{-1}\left(\frac{t}{t_\mrm{v}}\right)\,. \label{eq:nv}
\ee
Here $n_\mrm{b}$ is the bound fraction after violent relaxation and $t_\mrm{v}$ determines the timescale of this phase. The adopted values $n_\mrm{b}=0.1$ and $t_\mrm{v}=5\,\mrm{Myr}$ represent (roughly) the violent relaxation phase with a low global star formation efficiency of 15--20\%. 

The rate of change of the bound-star-number fraction due to cluster dissolution is given in our model in the following form:
\be
\frac{\d n_\mrm{D}}{\d t} = -c \left( \frac{M}{M_\mrm{br}} \right)^{(a_1-1)}\left[ 1 + \left( \frac{M}{M_\mrm{br}} \right)^{a_2} \right] n_\mrm{D}^{a_1}\,.
\label{eq:dnd}
\ee
with the boundary condition $n_\mrm{D}(t=0)=1$. Here, the rate is proportional to the power $a_1$ of the current value of $n_\mrm{D}$ and the constant, $c,$ determines the global rate of long-term decrease of this fraction. 

By construction, the rate is proportional to $M^{a_1-1}$ or $M^{a_1 + a_2-1}$ for initial masses below or above $M_\mrm{br}$, respectively. Having different lifetime scaling for low-number and high-number clusters allows the model to reproduce the observed range of cluster ages without creating an overabundance of very old high-number clusters.

The solution of Eq.~(\ref{eq:dnd}) can written in a compact form:
\begin{align}
n_\mrm{D}(t) &= \left(1 - \frac{t}{\tau(M)} \right)^{\textstyle\frac{1}{1-a_1}}\,,
\label{eq:nD} \\
\tau(M) &= \left\{c(1-a_1) \left( \frac{M}{M_\mrm{br}} \right)^{(a_1-1)}\left[ 1 + \left(\displaystyle \frac{M}{M_\mrm{br}} \right)^{a_2}\right]\right\}^{-1} \,.
\label{eq:lt}
\end{align}
For $a_1<1$, we have  $\tau(M)$ giving the lifetime of the clusters.

\begin{figure}[t!]
    \centering
    \includegraphics[width=0.9\hsize]{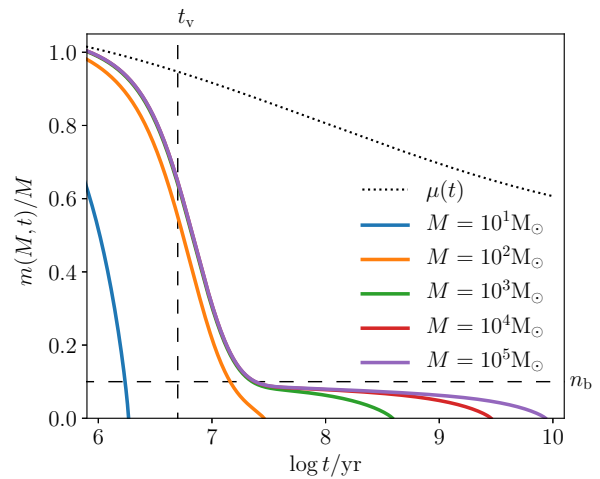}
    \caption{Evolution of the modelled mass fraction for clusters of different initial masses. The mass loss just from stellar evolution, $\mu (t)$, is shown in addition for comparison. The horizontal dashed line marks the bound fraction, $n_\mrm{b}$, after violent relaxation and the vertical dashed line marks the violent relaxation timescale, $t_\mrm{v}$.
    }
    \label{fig:modelMass}
\end{figure}

Next, we combine all three contributions to the cluster bound mass as a function of initial mass ($M$) and age ($t$):
\be \label{eq:mt}
m(M,t) = \mu(t)\, n_\mrm{v}(t)\, n_\mrm{D}(M,t) M\,.
\ee
The bound-mass fractions ($m(M,t)/{M}$) for some values of initial mass are shown in Fig.~\ref{fig:modelMass}. For clusters with an initial mass of $\log M/\msun > 2$, fast mass loss by violent relaxation is 
clearly separated from the long-term mass loss by stellar evolution and cluster dissolution in the tidal field. Low-mass clusters do not survive the violent relaxation phase.

\begin{figure*}[t]
   \centering
\includegraphics[width=0.525\hsize,clip=]{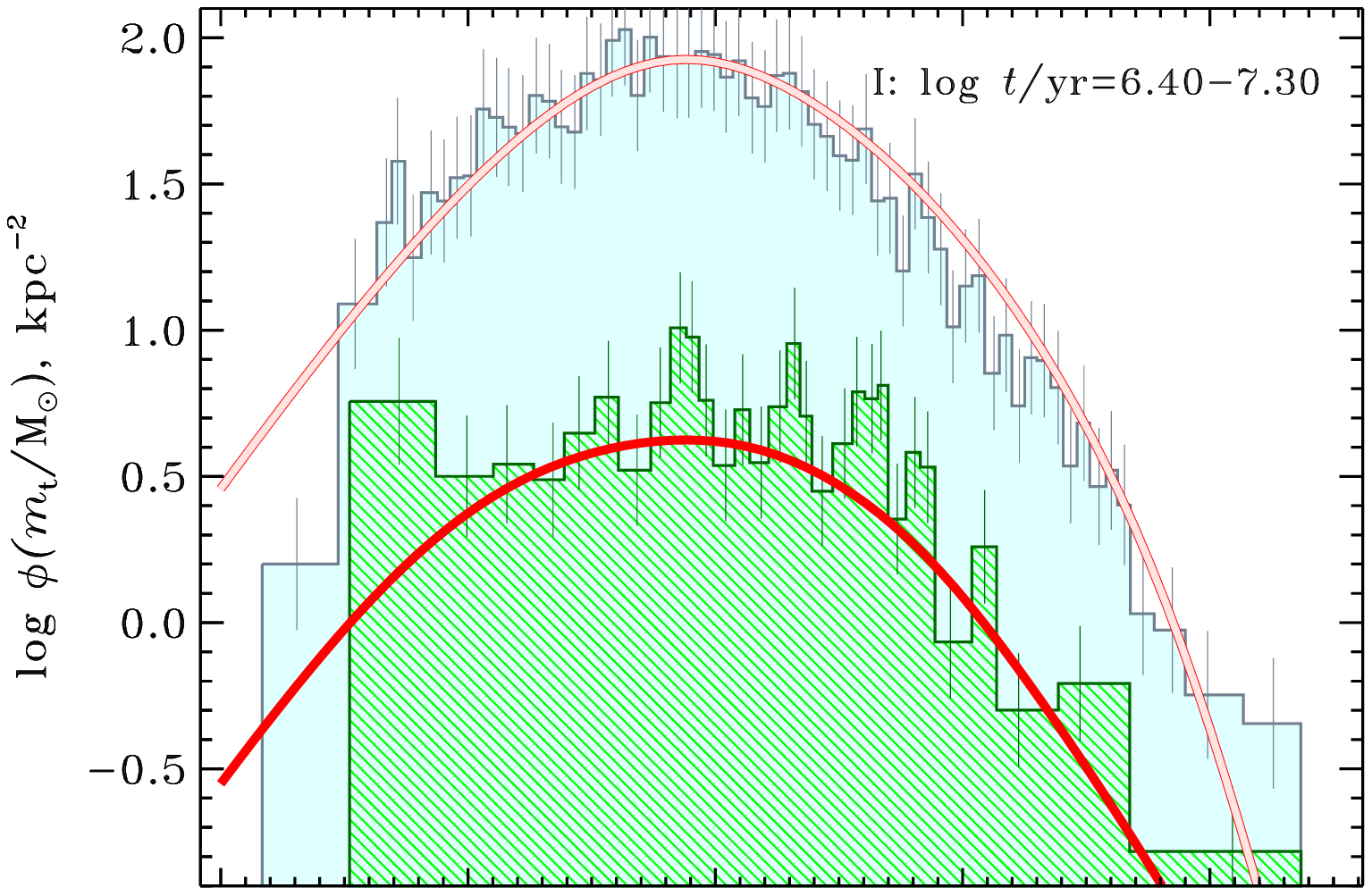}
\includegraphics[width=0.45\hsize,clip=]{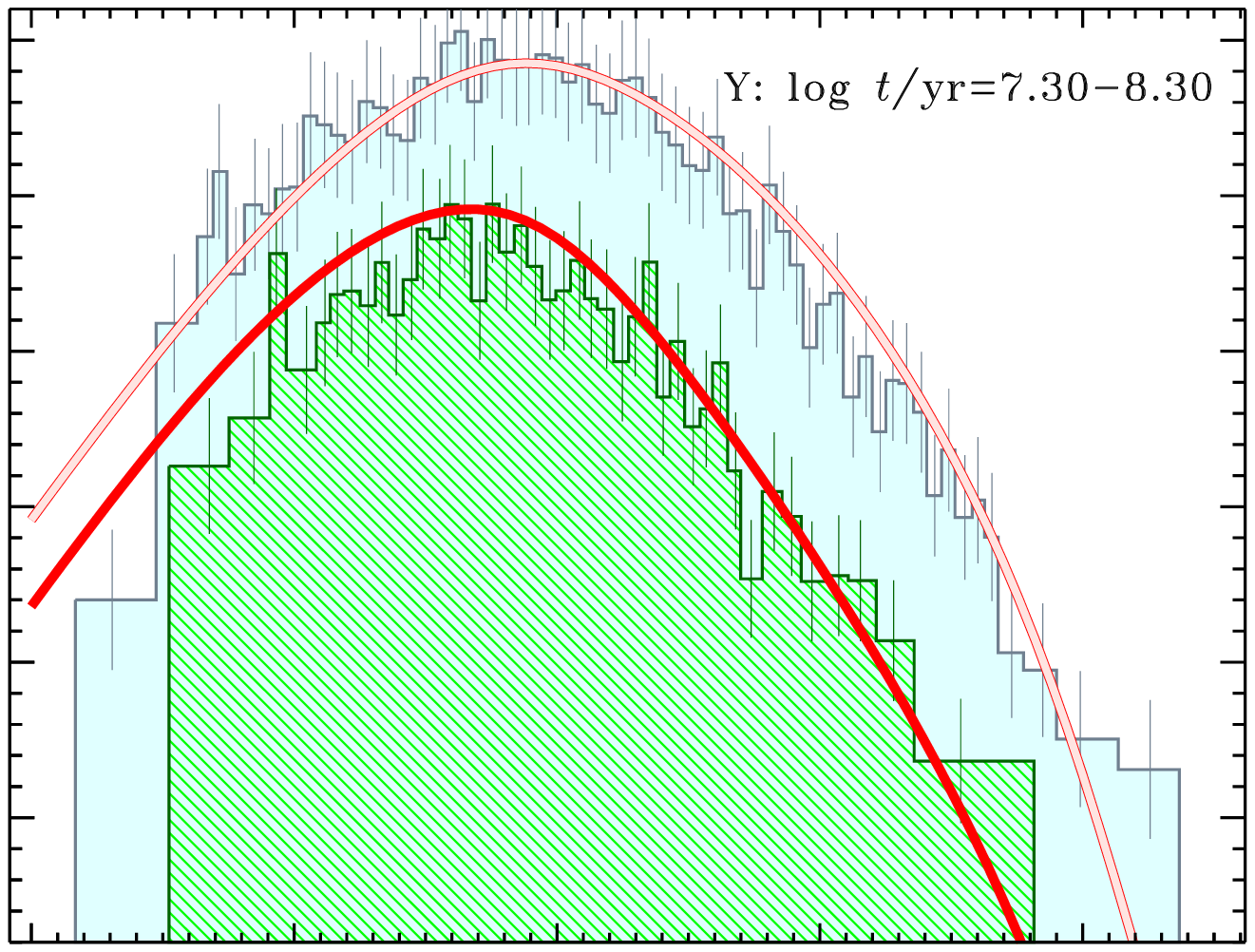}
\includegraphics[width=0.525\hsize,clip=]{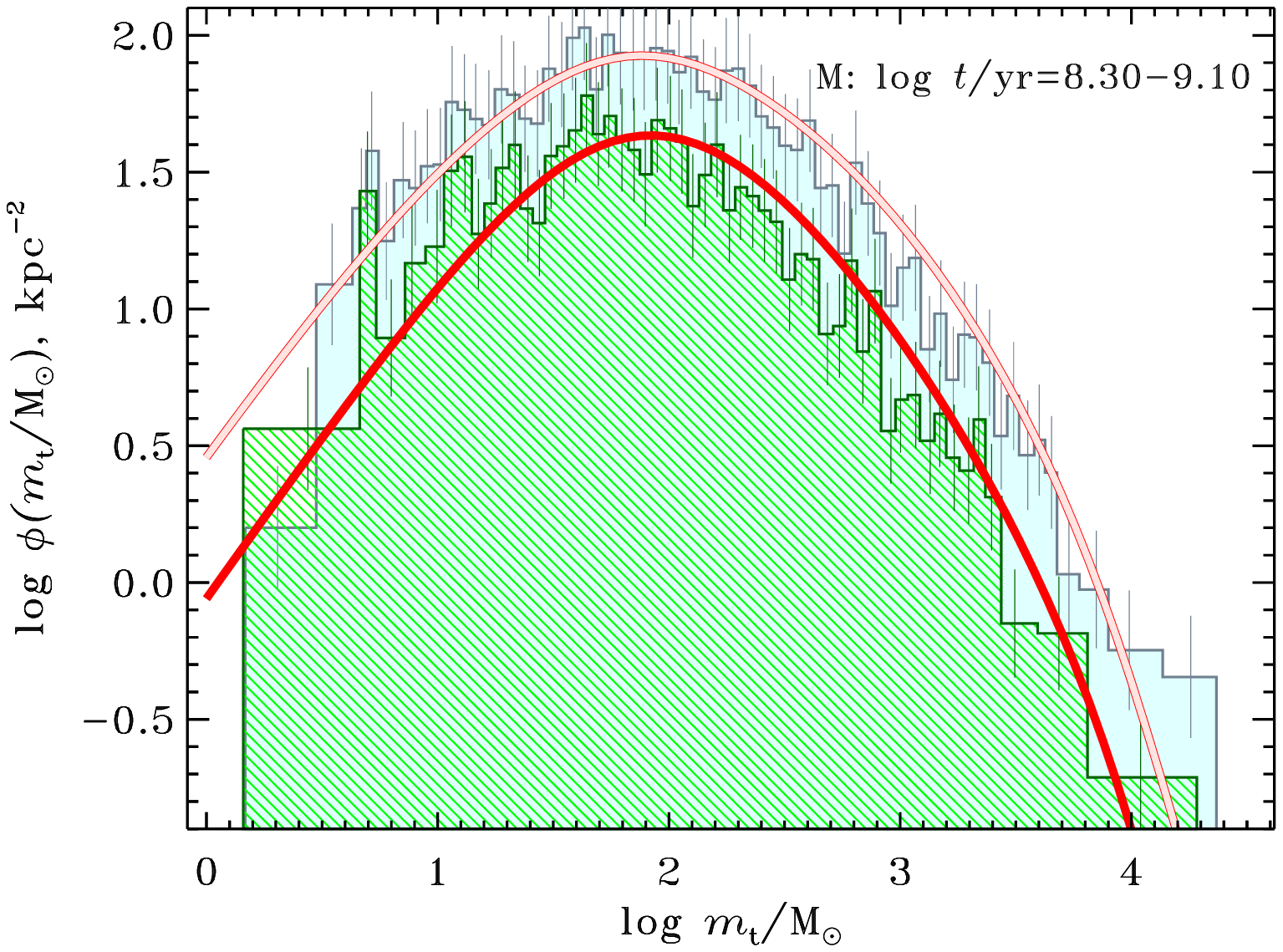}
\includegraphics[width=0.45\hsize,clip=]{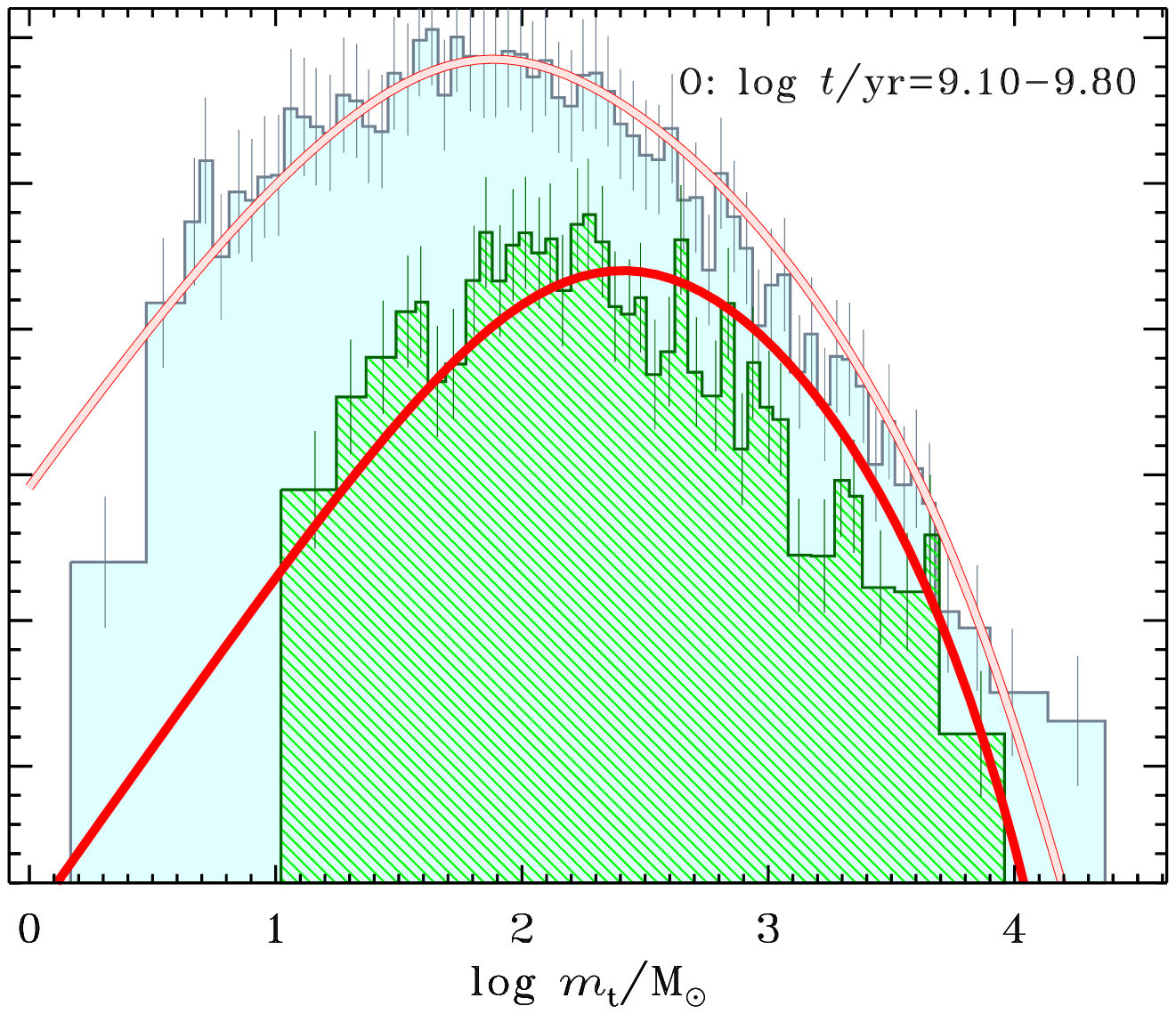}
\caption{Comparison of modelled (solid red curves) and observed mass distributions for the same age groups as shown in Fig.~\ref{fig:mf_iy} (green foreground histograms). The background histogram shows the GCMF constructed in Sect.~\ref{sec:clumf}. The hollow red curve is the modelled GCMF. Other designations are the same as in Fig.\ref{fig:mf_iy}.
}
\label{fig:modelCMF}
\end{figure*}

In Sect.~7.1, we assumed that the partial derivative ($\p m/\p M$) in Eq.~(\ref{eq:sigma2}) is positive for all $M$ and $t$. We check the constraints under which Eq.~\ref{eq:mt} fulfils this condition. With $n_{\rm D}$ given by Eqs.~\ref{eq:nD} and \ref{eq:lt}, the logarithmic derivative can be written in the form:
\be
 \frac{\p \log m}{\p \log M} =\frac{M}{m} \frac{\p m}{\p M} = \frac{\tau}{\tau-t} 
 \left\{1 - \frac{a_2}{1-a_1} \frac{t}{\tau}  \left[ \frac{M^{a_2}}{M^{a_2} + M^{a_2}_\mrm{br}}  \right] \right\}\,.
 \label{eq:dmdM}
\ee
Since for $a_1<1$ the age ($t$) is always smaller than $\tau(M)$ and the expression in square brackets can reach unity, the positiveness of the right-hand side of Eq.~\ref{eq:dmdM} requires:
\be
    a_1 + a_2 < 1\,,
    \label{eq:constrain}
\ee
which is fulfilled in our model.

\subsection{Comparison of the model and data}\label{sec:modr}

A comparison between the modelled and observed GCMF and CMFs of the different age groups is shown in Fig.~\ref{fig:modelCMF}.
The corresponding total cluster number surface densities ($\Sigma_{\rm N} $) are presented in Table~\ref{tab:model_densities} (with Eq. \ref{eq:Sigma_model} using the mass range of $\log m / \msun = 0-4.5$). 
The GCMF is very well reproduced by the model over the full mass range, except for the lowest and highest mass bins. The corresponding surface density of clusters is reproduced by construction.

An inspection of the CMFs of the different age groups provides more information about the quality of the model.
We find that the agreement among the observed and modelled CMFs across all age groups is good. 
There remain some minor systematic deviations of the model compared to the data in shape and normalisation of the CMFs. There is a deficit of very low mass clusters in the I-group (top-left panel of Fig. \ref{fig:modelCMF}), an overabundance of low-mass clusters in the Y-group (top-right panel of Fig. \ref{fig:modelCMF}),  
and a shift of the CMF to higher masses in the O-group (bottom-right panel of Fig. \ref{fig:modelCMF}). The model predicts a more peaked distribution of low-mass clusters at 50~Myr, leading to an excess of clusters in the Y-group and a deficit in the I- and M-group (Table~\ref{tab:model_densities}).

\begin{table}[t]
    \tabcolsep=6pt
    \caption{Total surface number densities of clusters ($\Sigma_{\rm N}$, Eq. \ref{eq:Sigma_model}) for each age group. }
    \label{tab:model_densities}
    \centering
    \begin{tabular}{c c c}  
        \hline
        \hline
        \vgap
         Age group & Observed, & Modelled,\\
         & $\mrm{kpc}^{-2}$  & $\mrm{kpc}^{-2}$ \\
         \vgap
         \hline
         \vgap
         Initial-Age (I) & 13.3 & 8.0 \\
         Young (Y) & 31.9 & 41.2 \\
         Medium-Age (M) & 65.2 & 61.8 \\
         Old (O) & 23.8 & 22.9 \\
         \vgap
\hline
\vgap
         General (G) & 136.9 & 135.7 \\
         \vgap
         \hline
    \end{tabular}
\end{table}

\begin{figure}[t!]
    \centering
    \includegraphics[width=0.9\hsize]{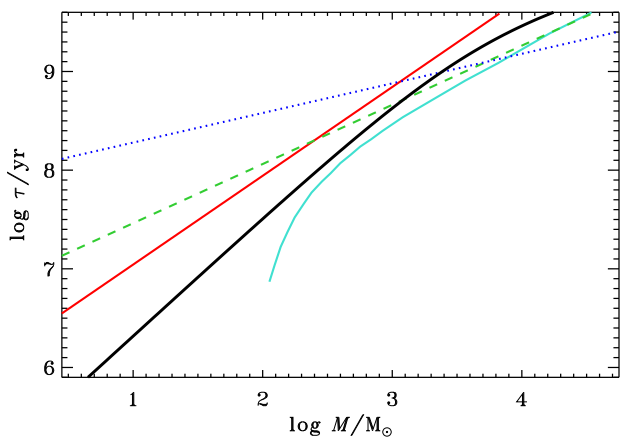}
    \caption{Lifetime-mass relation (Eq.~\ref{eq:lt}) of the present model (black) compared  to data of \citet{lamgi06} for 100 $M_\sun$-cluster remnant (light blue) and the parametrisation covering the results based on N-body calculations of \citet{ernstea15} used in  \citetalias{mwscage}. The red line corresponds to underfilling clusters, the green dashed line represents Roche volume filling models, and the blue dotted line is for overfilling models.
    }
    \label{fig:lt}
\end{figure}

Our resulting cluster lifetime (see Eq.~\ref{eq:lt} and Fig.~\ref{fig:lt}) follows a broken power law, which is steeper at the low-mass end and shallower at the high-mass end. 
For high-mass clusters, $M>1000\,\msun$, our lifetimes till complete dissolution are larger than the dissolution time in \citet{shukirea18}. We note that they use the cluster mass after violent relaxation (corresponding to $n_\mrm{b}M$ in this work) and define the dissolution time as the time when the cluster mass falls below 100\,$\msun$. The low-mass regime is not covered in their investigation.
Figure \ref{fig:lt} also shows the cluster lifetimes from \citet{ernstea15}, which we used in \citetalias{mwscage}. In this work, the lifetime falls between the line for underfilling clusters (red line) and that based on the data of \citet{lamgi06} for 100 $M_\sun$-cluster remnants (solid light blue line).

The mean mass of the CIMF in the mass range $\log M/\msun \geq$ 0.3 is 505~$\msun$. Combined with the adopted CFR, we find a formation rate in the mass of 409~$\msun\,\mrm{kpc}^{-2}\,\mrm{Myr}^{-1}$. A comparison with the present-day thin disc star formation rate of \citet{sysjus21} leads to a fraction of 30\% of field stars formed in clusters including that 90~percent of stars that were lost during the violent relaxation phase. 
This value is somewhat smaller but still consistent, due to the large uncertainties, with a fraction of up to 40\% found in \citet{fuma} and \citet{roeserea10}. 

\section{Summary and conclusions}\label{sec:conc}

This paper reports the results of the second part of our study on the history of the formation and evolution of Galactic star clusters based on the MWSC catalogue. In the first part, published in \citetalias{mwscage}, we constructed the global cluster age distribution. We compared the observations with the outcome of a model of the star cluster population of the MW disc. This model included a power law for the lifetime-mass relation, an exponentially declining CFR, and a broken-power-law CIMF. For three different lifetime-mass relations, we found similar good fits for the cluster age distribution but with very different pairs of CFR and CIMF. This degeneracy could not be resolved without additional information about cluster masses.

In the current paper, we determine the tidal masses of the star clusters and derive the CMFs for different age groups to resolve the ambiguity between CFR, CIMF, and cluster mass evolution. We start with the same sample of MWSC star clusters as in \citetalias{mwscage}. It includes 3063 open clusters with homogeneous determinations of spatial-kinematic and astrophysical parameters based on combined kinematic and photometric cluster membership criteria. For 98\,\% of the clusters, the structural King parameters (including tidal radii with a typical accuracy of 30\%) are determined. For the most massive clusters, the tidal radius varies from 15~pc near the MW centre to more than 50~pc in the periphery. The sample occupies an extended disc area between Galactocentric radii of 2--20~kpc. As our analysis has shown, the sample is complete for the brightest clusters inside the heliocentric cylinder with a radius of up to 5 kpc. Due to the nature of the MWSC, which is close to a magnitude-limited survey, the radius of completeness depends on the brightness of the clusters and for the faintest objects, it is as small as about 1 kpc. In total,  inside the magnitude-dependent completeness limits, there are 2227 MWSC clusters of different ages and brightness.

Equating the tidal radius to the Jacobi radius, we derive a tidal mass ($m_\mrm{t}$) from the balance between Milky Way's tidal field and cluster's gravity. The typical accuracy of its determination is 70\%. The tidal masses span over $\log m_\mrm{t}/\msun = 0.2-4.4$. The upper limit of this range remains constant throughout all Galactocentric distances covered by
the MWSC, except for a dozen of the most massive clusters in the Galactic centre (Fig.\,\ref{fig:rmt_rg}). The latter were excluded from the analysis: we believe that they suffer from an underestimation of interstellar extinction due to the patched cloud-dominated structure of a dust layer in the Galactic centre.

Using the tidal masses of all clusters within their magnitude-dependent completeness limits, we first built the classic GCMF. It has a bell-like shape and has a power-law dependence for $\log m_\mrm{t}/\msun>2.3$, as has been observed for clusters in other galaxies. However, the origin of the apparent maximum is different. In external galaxies, its position depends on the distance to the galaxy, which likely points to data incompleteness below the observation limit. The position of the GCMF maximum obtained here is a consequence of the mass distribution of newly formed clusters and their subsequent evolution. Uncertainties in tidal radii propagated to tidal masses dominate the errors in the GCMF values. The linear best fit of the high-mass slope in the mass range $\log m_\mrm{t}/\msun=$ 2.3--4.4 turned out to be $x=$ 1.14$\pm$0.07. This result agrees with previous Milky Way studies based on our COCD results \citep{lamea,fuma}.

Then, we analysed variations of the CMF with age and location. To approach the age variation, we divided the general sample into four age sub-samples, built the corresponding CMFs, and determined their individual parameters. We find that similar to the GCMF and independent of the age range, they consist of two qualitatively different segments. In the low-mass range, $\log m_\mrm{t}/\msun \lesssim 2.0$, the CMFs grow with mass. In the high-mass end, $\log m_\mrm{t}/\msun \gtrsim 2.0$, they decrease with mass. The slopes  do not change significantly with age, except for the low-mass slope of the initial-age (I) sub-sample; in this case, we obtained a nearly flat distribution at $\log m_\mrm{t}/\msun \lesssim 2.3$. The I-sub-sample contains clusters of age between 2.5--20 Myr. On this timescale, violent relaxation is the dominant mechanism of mass function evolution. 

To study the dependence of the CMF on spacial location, we compare the CMFs for sub-samples of inner and outer clusters observed in Galactocentric zones with $R=$ 4.2--8.1~kpc and 8.9--13.5~kpc. Applying our standard procedure, we find that the shapes of CMFs are consistent in general with the GCMF centred at the position of the Sun, but there are small systematic vertical shifts relative to the GCMF. The inner CMF shows an overabundance, meantime the outer CMF is underabundant compared to the GCMF.
The differences are larger for higher cluster masses, which cover a larger Galactocentric radius range. This variation of the cluster number density with Galactocentric radius is consistent with the surface density profile of the Galactic disc exponentially decreasing.
Since other inhomogeneities such as spiral arms may also bias the observed CMF, a more detailed analysis of the cluster distribution is necessary for deriving the radial scale length, which is beyond the scope of this paper.

In \citetalias{mwscage}, we provided a simple analytic model of cluster formation and evolution to reproduce the global age distribution. In the current paper, using the extended data set including the CMFs in four age groups, we can separate the effects of the CIMF and the cluster mass evolution by using the bound mass ($m_\mrm{t}$) instead of the lifetime as a function of the initial mass. The CMFs in the different age groups depend on the CIMF, the CFR, and the cluster bound-mass function. We argue that the differences in the CMFs of the two young populations (I-group at 2.5--20 Myr and Y-group at 20--200 Myr) can be understood by accounting for a strongly enhanced mass loss in the first 20 Myr of cluster evolution. This agrees well with the cluster formation and evolution models of \citet{shukirea17,shukirea18} based on a concentrated star formation efficiency. Our result supports the models with low global star formation efficiency of approximately 16\%, where 90\% of the stars are lost in the violent relaxation phase due to the supervirial state after gas expulsion.

Considering an analysis of the four age groups division, it is impossible to derive details of the CFR because of the strong degeneracy with the cluster mass loss. For instance, a time-dependent CFR leads to vertical shifts of the model lines in Fig.~\ref{fig:modelCMF} without changing its profiles. As is seen from the figure, this can improve the CMF fit of the youngest I-group only. Here, an additional recent cluster formation event adding about 15\% young clusters would result in a better fit of the CMF. An appropriate star formation event was recently found  in A-type field stars in the solar neighbourhood \citep{sysjus21}. We also tested the effect of an exponentially declining CFR according to \citet{aumebin09} with a decay timescale of 8.5\,Gyr. With a slight adaption of the mass loss parameters, a similarly good fit of the CMFs was found, which is an indication of the degeneracy between the CFR and the cluster mass loss.

Our mean CMF corresponds to a cluster formation density of about 0.4~$\msun\,\mrm{pc}^{-2}\mrm{Gyr}^{-1}$, including the mass loss during the violent relaxation phase. This gives a cluster contribution of 30\% to the stellar content of the thin disc \citep{sysjus21}, which is lower than the estimate derived by \citet{fuma} and \citet{roeserea10}. In addition, clusters that dissolve completely after gas removal, a phenomenon that is sometimes referred to as 'infant mortality', are not included in this fraction. 

In future work, we will investigate the full 2D cluster distribution $\sigma(m,t)$  as a function of current mass and age, including a large parameter study and more detailed fits of the cluster mass evolution. 

\begin{acknowledgements}
We thank the anonymous referee for his detailed comments, which helped to improve the paper. This study was partly supported by the
Russian Foundation of Basic Research grant  20-52-12009. We would like to acknowledge the previous work with our colleagues N.V.~Kharchenko, P.~Berczik and M.~Ishchenko, who participated in the discussions at the beginning of this work. 
We acknowledge the use of the Simbad database, the VizieR Catalogue Service and other services operated at the CDS, France and the WEBDA facility operated at the Department of Theoretical Physics and Astrophysics of the Masaryk University. 
\end{acknowledgements}

\bibliographystyle{aa}
\bibliography{clubib}

\end{document}